\documentclass[10pt,final,doublecolumn]{IEEEtran}
\usepackage{overpic}
\usepackage{algorithmic}
\usepackage{amsmath}
\usepackage{amssymb}
\usepackage{amsfonts}
\usepackage{latexsym}
\usepackage{graphicx}
\usepackage{bbding}
\usepackage{enumerate}
\usepackage{multicol}
\usepackage{color}
\usepackage{slashbox}
\usepackage{setspace}
\usepackage{subfigure}
\usepackage[ruled,linesnumbered]{algorithm2e}

\IEEEoverridecommandlockouts
\allowdisplaybreaks[4]

\begin{document}

\title{A Dimension Reduction-Based Joint Activity Detection and Channel Estimation Algorithm for Massive Access}
\author{
Xiaodan Shao, Xiaoming Chen, and Rundong Jia
\thanks{Xiaodan Shao ({\tt shaoxiaodan@zju.edu.cn}), Xiaoming Chen ({\tt chen\_xiaoming@zju.edu.cn}) and Rundong Jia ({\tt jia\_rundong@zju.edu.cn}) are with the College of Information Science and Electronic Engineering, Zhejiang University, Hangzhou 310027, China.}}
\maketitle

\begin{abstract}
Grant-free random access is a promising protocol to support massive access in beyond fifth-generation (B5G) cellular Internet-of-Things (IoT) with sporadic traffic. Specifically, in each coherence interval, the base station (BS) performs joint activity detection and channel estimation (JADCE) before data transmission. Due to the deployment of a large-scale antennas array and the existence of a huge number of IoT devices, JADCE usually has high computational complexity and needs long pilot sequences. To solve these challenges, this paper proposes a dimension reduction method, which projects the original device state matrix to a low-dimensional space by exploiting its sparse and low-rank structure. Then, we develop an optimized design framework with a coupled full column rank constraint for JADCE to reduce the size of the search space. However, the resulting problem is non-convex and highly intractable, for which the conventional convex relaxation approaches are inapplicable. To this end, we propose a logarithmic smoothing method for the non-smoothed objective function and transform the interested matrix to a positive semidefinite matrix, followed by giving a Riemannian trust-region algorithm to solve the problem in complex field. Simulation results show that the proposed algorithm is efficient to a large-scale JADCE problem and requires shorter pilot sequences than the state-of-art algorithms which only exploit the sparsity of device state matrix.
\end{abstract}

\begin{IEEEkeywords}
B5G, grant-free, activity detection, channel estimation, massive connectivity, Riemannian optimization.
\end{IEEEkeywords}

\section{Introduction}
The widespread use of IoT in smart traffic, smart city, smart health care and factory automation spurs an explosive growth of IoT devices in the past years \cite{mtc,mtc1}. In general, IoT has two prominent characteristics. First, the number of IoT devices is very large and increasingly grows \cite{IoT1}. It is predicted that the number of IoT devices will reach hundreds of billions in 2030. Second, the data of IoT devices has a sporadic traffic pattern \cite{IoT2}. In other words, only a small fraction of huge number of potential devices are active at a given time. The grant-based random access adopted in 5G Narrowband Internet of Things (NB-IoT) may lead to a high latency with the further development of IoT. To this end, the grant-free random access protocol that allows massive IoT devices to simultaneously access wireless network without a grant is considered in future B5G cellular IoT \cite{free}. In this context, grant-free random access receives considerable attentions recently. In \cite{unsource}, the authors proposed the unsourced massive random access. Then based on this work, \cite{unsource1} extended the massive unsourced random access to the case where the base station has a massive number of antennas and combined a grant-free non-coherent massive  multiple-input multiple-output (MIMO) activity detection scheme.

For the grant-free random access protocol, each device is pre-assigned with a unique pilot sequence. The BS detects the devices' states by processing the received pilot sequences. Meanwhile, the BS is able to obtain channel state information (CSI), which is used for decoding the uplink signals and precoding the downlink signals in a time division duplex (TDD) mode. Hence, the algorithm design for grant-free random access has a vital impact on the performance of B5G cellular IoT. Since inactive devices do not transmit their pilot sequences, a device state matrix containing device and channel state information is typically sparse, which by nature introduces a compressed sensing (CS) problem \cite{gaozhen1}. Several CS methods were firstly employed to detect devices¡¯ activity and their data by assuming that devices' CSI was known in advance \cite{datad1}-\cite{datad3}. Afterward, in \cite{AMPO}-\cite{gaozhen2}, a joint activity detection and channel estimation via an approximate message passing (AMP) algorithm was adopted for a single-cell system. \cite{AMPO} proposed an AMP algorithm design that exploits the statistics of the wireless channel and provides an analytical characterization of the probabilities of false alarm and missed detection by using the state evolution. \cite{AMPi} showed that in the asymptotic massive MIMO regime, both the missed device detection and the false alarm probabilities for activity detection can always be made to go to zero by utilizing AMP algorithm. For the payload data containing only a few bits, \cite{AMPii} devised a new non-coherent transmission scheme for  massive machine-type communications (mMTC) and specifically for grant-free random access by leveraging elements from the approximate message passing algorithm. The authors in \cite{multicell,multicell1} studied the multi-cell user activity detection problem for massive connectivity and characterized the performance for both the massive MIMO and the cooperative MIMO architectures. The device activity detection of a cloud-radio access network (C-RAN) also has been studied in \cite{cran,cran1}. Specifically, the authors in \cite{cran} investigated the impact of fronthaul capacity limitations in a C-RAN architecture on the functions of random access and active device identification in the massive access scenario. The authors in \cite{cran1} proposed Bayesian compressive sensing-based algorithm and try to exploit the prior channel information of path loss effects and the chunk sparsity structure to solve the problem of JADCE in uplink C-RAN. In \cite{OMP1} and \cite{OMP}, a greedy CS algorithm based on an orthogonal matching pursuit was designed for sparse signal recovery from random measurements. In \cite{EP}, the authors proposed an expectation propagation based JADCE algorithm for massive access networks with the aid of channel prior-information.

These algorithms exploiting the sparsity structure of the device state matrix have good detection and estimation performance, but face several practically challenging issues. Firstly, the deployment of a large-scale antenna array at the BS and the existence of a massive number of IoT devices form a large-dimensional device state matrix result in prohibitively computational complexity. Secondly, the concurrent transmission of pilot sequences leads to severe interference. In order to guarantee the accuracy of detection and estimation, long pilot sequences have to be utilized, which decreases the efficiency of information transmission. Some of the recent works to solve these problems include \cite{scaling,scaling1}, where the authors proposed a low-complexity covariance based approach for device activity detection, whose solution depends on the received signal through certain covariance matrix only, and such an approach can exploit the multiple BS antennas more effectively. Then, the authors in \cite{covarichen} provide an accurate performance analysis, in terms of the probabilities of false alarm and missed detection, for the joint device activity detection and data decoding scheme using the covariance based approach.

In fact, since the traffic is sporadic, the device state matrix is typically low-rank. Motivated by this observation, the joint sparse and low-rank structure can be exploited to reduce the computational complexity and the training overhead. Note that the low-rank problem is in general NP-hard, convex relaxations have to be utilized to solve the CS problem based on the nuclear norm. For examples, the work \cite{convex1} proposed an iterative algorithm by replacing the rank function with the nuclear norm. Yet, the performance of such an iterative algorithm heavily depends on the initiation value. Furthermore, the work \cite{convex2} proposed a semidefinite programming (SDP) solver via nuclear norm relaxation. However, the computational and memory requirements for solving an SDP problem limit its applicability in moderate to high-dimensional data problem. In short, the nuclear norm based convex relaxation approaches fail to well incorporate the fixed-rank matrices for sparse signal recovery problem due to the poor structures.

For achieving grant-free random access in B5G cellular IoT, it is required to design a scalable computational and efficient algorithm to robustly detect device activity and estimate channel information based on the sparse and low-rank characteristics of device state matrix. In this context, this paper first provides a dimension deduction method for the device state matrix, such that JADCE is reduced to a low-dimensional optimization problem with a full column rank constraint. To decrease the search space of the JADCE problem with the nonconvex fixed-rank constraint, this paper proposes a Riemannian trust region algorithm by making use of the non-compact Stiefel manifold of fixed-rank matrices in complex field. Especially, by harnessing the second order information on Riemannian manifold \cite{rieman}-\cite{pa}, the proposed algorithm achieves a superlinear convergence rate and converges to first-order and second-order KKT points on manifolds from arbitrary initial points. As compared to the AMP based approach, the proposed algorithm achieves lower detection error probabilities. Moreover, the proposed algorithm does not require the knowledge of large-scale fading coefficients of the channel, statistics of the channel vectors and number of active users per coherence time, which are quite critical for deriving AMP but are difficult to have in practice due to the sporadic traffic. Compared with coordinate-wise descend activity detection algorithm including Multiple Measurement Vector (MMV), Non-Negative Least Squares (NNLS) and Maximum Likelihood (ML) estimators \cite{scaling}, the activity detection performance of the proposed algorithm is better than MMV estimator, NNLS estimator, and worse than ML estimator. More importantly, the computation complexity of the proposed manifold-related operations for solving JADCE problem does not increase by increasing the number of BS antennas, which can effectively reduce computational complexity when the number of BS antennas is huge, as is often the case for mMTC. The contributions of this paper are as follows:

\begin{enumerate}

\item This paper provides a simple rank estimation method for the interested device state matrix based on the received signal, and then designs a dimension reduction method for JADCE, which projects the original problem to a low-dimensional space where the interested device state matrix has full column rank as estimated.

\item This paper develops a novel JADCE framework with coupled full column rank constraint thereby reducing the size of the search space, and further designs a logarithmic smoothing method for the resulting objective function.

\item This paper proposes a novel Riemannian trust region algorithm for JADCE, and it is found that the proposed algorithm is robust to the estimation error of the rank of the device state matrix.

\end{enumerate}

The rest of this paper is organized as follows: Section II gives a brief introduction of B5G cellular IoT in a sporadic device activity pattern scenario. In Section III, we provide a dimension deduction method and a rank estimation method. Based on the both methods, we present a JADCE framework. In Section IV, we propose a Riemannian optimization algorithm for the design of JADCE. Section V provides extensive simulation results to validate the effectiveness of the proposed algorithm. Finally, Section VI concludes the whole paper.

\emph{Notations}: We use bold letters to denote matrices or vectors, non-bold letters to denote scalars, $(\cdot)^H$ to denote conjugate transpose, $\mathbf{I}_N$ to denote the identity matrix of order $N$, $\mathbb{C}^{A\times B}$ to denote the space of complex matrices of size $A\times B$, and $\mathbf{e} \sim\mathcal{CN}(\mathbf{0},\sigma^2\mathbf{I)}$ to denote that each element in $\mathbf{e}$ follows the independent and identically distributed (i.i.d.) Gaussian distribution with zero mean and variance $\sigma^2$, $\|\cdot\|_F$ to denote the Frobenius norm of a matrix, $\|\mathbf{a}\|$ to denote the $l_2$ norm of a vector $\mathbf{a}$, $|\cdot|_c$ to denote the cardinality of a set, $\mathbf{A}[n]$ to denote the $n$th row of matrix $\mathbf{A}$. $\left \| \mathbf{A} \right \|_{2,1}$ is the $l_{21}$ norm defined as the sum of $\mathbf{A}[n]$.

\section{System Model}
\begin{figure}[t]
  \centering
\includegraphics [width=0.45\textwidth] {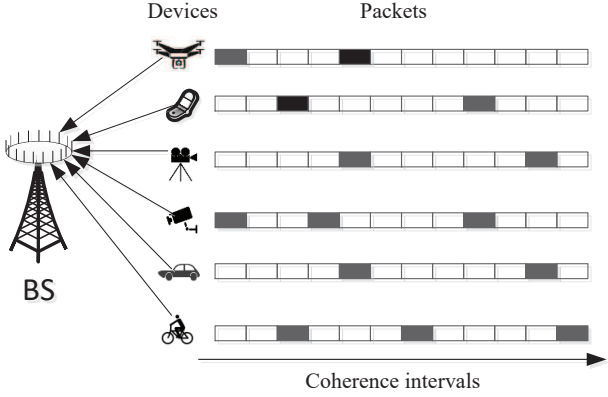}
\caption{A B5G cellular IoT network with sporadic traffic devices.}
\label{MUD}
\end{figure}

We consider a B5G cellular IoT network, where a BS equipped with $M$ antennas serves $N$ single-antenna IoT devices. Note that in B5G cellular IoT, the number $N$ of IoT devices is usually very large. However, due to the burst characteristic of IoT applications, only a fraction of IoT devices are active in a coherence interval, as shown in Fig. \ref{MUD}. We use a signal support $\mathcal{K}$ to denote the set of active devices in a certain coherence interval and $K=\left|\mathcal{K}\right|_c$ to denote the number of active devices. For convenience, we define $\chi_n$ as the activity indicator with ${\chi_n} = 1$ if the $n$th device is active, otherwise, ${\chi_n} = 0$. Let $\varepsilon_n$ represent the active probability of the $n$th device, then we have
\begin{equation}
\label{eqinte2,r}
\left\{ \begin{array}{l}
\Pr ({\chi_n} = 1) = \varepsilon_n,~n\in\mathcal{K} \\
\Pr ({\chi _n} = 0) = 1 - \varepsilon_n,~\textrm{otherwise}
\end{array} \right.
\end{equation}

A grant-free random access protocol is adopted, such that a massive number of IoT devices with sporadic data traffic can access the B5G wireless network instantaneously without a grant to transmit or a prior scheduling assignment. The $n$th IoT device is assigned a unique i.i.d. complex Gaussian distribution pilot sequence $\mathbf{a}_{n}=[a_{n,1},\cdots ,a_{n,L}]^T\in\mathbb{C}^{L\times1}$ with unit norm, where $L$ is the length of pilot sequences. At the beginning of a coherence interval, active IoT devices simultaneously send their pilot sequences over the uplink channels for joint activity detection and channel estimation (JADCE) at the BS. Thus, the received signal $\mathbf{Y}\in \mathbb{C}^{L\times M}$ at the BS can be expressed as
\begin{eqnarray}
\label{eqinter5}
\mathbf{Y}
&=&\sum_{n=1}^{N}\chi_n\sqrt{\varsigma_n}\mathbf{a}_n\mathbf{h}_n^T+\mathbf{E}\nonumber\\
&=&\mathbf{AX+E},
\end{eqnarray}
where $\mathbf{h}_n=\sqrt{\vartheta }\tilde{\mathbf{h}}_n$ denotes the $M$-dimensional channel vector from the $n$th device to the BS, which composed of large-scale fading $\vartheta$ and small-scale fading $\tilde{\mathbf{h}}_n$. The channels maintain constant in a coherence interval and independently fade over coherence intervals. The $\mathbf{E}=[\mathbf{e}_1,\cdots ,\mathbf{e}_M]\in \mathbb{C}^{L\times M}$ with $\mathbf{e}_m\sim \mathcal{CN}(\mathbf{0},\sigma^2\mathbf{I)}$ is an additive white Gaussian noise (AWGN) matrix at the BS, where $\sigma^2$ is the variance of noise, and $\varsigma_n=Lp_n$ is the transmit energy with $p_n$ being the pilot transmit power at the $n$th device. For simplicity of notation, we define $\mathbf{X}=[\mathbf{x}_1,...,\mathbf{x}_N]^T\in\mathbb{C}^{N\times M}$ with $\mathbf{x}_n={\chi}_n\sqrt{\varsigma_n}\mathbf{h}_n\in\mathbb{C}^{M\times1}$ as a device state matrix and $\mathbf{A}=[\mathbf{a}_1,...,\mathbf{a}_N]\in\mathbb{C}^{L\times N}$ as a pilot sequence matrix.

Based on the received signal $\mathbf{Y}$, the BS recovers $\mathbf{X}$ for JADCE. Specifically, for a recovered $\mathbf{X}$, the $n$th device is judged as active if its elements meet a given condition, and the corresponding CSI is given by $\mathbf{x}_n/\sqrt{\varsigma_n}$. Since a large portion of IoT devices are inactive, namely $\chi_n=0$, $\mathbf{X}$ is a sparse matrix. Moreover, $\mathbf{X}$ is low-rank due to sporadic traffic and large BS antenna array. In what follows, we design a low-complexity JADCE algorithm by exploiting the sparse and low-rank structure of $\mathbf{X}$.

\section{Design of Joint Activity Detection and Channel Estimation}
In this section, we focus on the design of a JADCE algorithm, which can be formulated as a problem of recovering the device state matrix $\mathbf{X}$ from the received signal $\mathbf{Y}$. In general, the recovery of a sparse matrix $\mathbf{X}$ from measurements $\mathbf{Y}$ in (\ref{eqinter5}) is a typical multiple measurement vector problem in compressed sensing \cite{music}. A quite well-known technique for recovering the row-sparse matrix $\mathbf{X}$ in the  multiple measurement vector setting is the $l_{21}$ norm least square method as below:
\begin{equation}\label{mmv}
  \mathop \text{argmin}\limits_{\mathbf{X}\in \mathbb{C}^{N\times M}}\left \| \mathbf{X} \right \|_{2,1}+\frac{\zeta}{2} \left \| \mathbf{AX}-\mathbf{Y} \right \|_F^2,
\end{equation}
where $\zeta$ is a fixed penalty parameter. The main drawback of a MMV problem above is its high computational complexity due to the high dimensions of $\mathbf{A}$ and $\mathbf{X}$. Moreover, the accuracy of recovered $\mathbf{X}$ is dependent on the length $L$ of pilot sequences. However, a long pilot sequence leads to a short duration for information transmission. To solve these problems, we propose a dimension reduction-based JADCE algorithm.

\subsection{Dimension Reduction}
Prior to designing the JADCE algorithm, we first provide a dimension-reduced equivalent model for (\ref{mmv}). Since the device state matrix $\mathbf{X}$ is row-sparse in the presence of sporadic traffic, it is typically low-rank, namely $r^e=\text{rank}(\mathbf{X})\ll \text{min}\{M,N\}$. Based on such a characteristic, we have the following theorem:

\textbf{Theorem 1}: Partition the received signal $\mathbf{Y}$ into a signal space and its null spaces by singular value decomposition (SVD), namely $\mathbf{Y}=\mathbf{S}_{sd}\mathbf{V}_{sd}\mathbf{D}_{sd}^T$. Let $\mathbf{V}=\mathbf{S}_{r^e}\mathbf{V}_{r^e}$, where $\mathbf{S}_{r^e}$ is the first $r^e$ columns of $\mathbf{S}_{sd}$ and $\mathbf{V}_{r^e}$ is a square matrix consisting of the first $r^e$ rows and the first $r^e$ columns of $\mathbf{V}_{sd}$. Moreover, let $\mathbf{U}$ be the first $r^e$ rows of $\mathbf{D}_{sd}^T$. Then, we can construct a signal space as
\begin{equation}\label{signalspace}
\mathbf{VU}=\mathbf{A}\mathbf{X}+\mathbf{E}_{X},
\end{equation}
where $\mathbf{E}_{X}$ is the noise incorporated in the signal space, $\mathbf{V}\in \mathbb{C}^{L\times r^e}$ with $\text{rank}(\mathbf{V}) = r^e$, $\mathbf{U}\in \mathbb{C}^{r^e\times M}$ with $\mathbf{UU}^H=\mathbf{I}$. We emphasize that the result (\ref{signalspace}) is a high SNR characterization of the received data. Such a high-SNR approximation is a reasonable assumption in B5G cellular IoT with massive connectivity due to the interference limited. Based on the above signal space,

\begin{enumerate}

\item
we can construct an equivalent form to the original problem (\ref{eqinter5}) as
\begin{equation}\label{samd}
  \mathbf{V}=\mathbf{AS}+\mathbf{E}_{S},
\end{equation}
with $\mathbf{S}=\mathbf{X}\mathbf{U}^H$ and $\mathbf{E}_{S}=\mathbf{E}_{X}\mathbf{U}^H$. If $L\geq \mathcal{O}(K\ln(N/K)+Kr^e)$, the equivalent problem
\begin{equation}\label{eq3}
  \mathop \text{argmin}\limits_{\mathbf{S}\in \mathbb{C}^{N\times r^e}}\left \| \mathbf{S} \right \|_{2,1}+\frac{\zeta}{2} \left \| \mathbf{AS}-\mathbf{V} \right \|_F^2,
\end{equation}
can stably reconstruct a $K$ row-sparse (having at most $K$'s non-zero rows) matrix $\mathbf{S}\in \mathbb{C}^{N\times r^e}$ with $\text{rank}(\mathbf{S})=r^e$.

\item
The originally concerned device state matrix ${\mathbf{X}}$ can be obtained by letting ${\mathbf{X}}=\mathbf{S}\mathbf{U}$.
\end{enumerate}

\begin{IEEEproof}
In order to project $\mathbf{Y}$ into a low-dimensional subspace while minimizing the loss of variability of $\mathbf{Y}$, the rank $r^e$ of the device state matrix $\mathbf{X}$ under the model (\ref{eqinter5}) needs to be known in advance. We will give a rank estimation method based on the received signal $\mathbf{Y}$ in Section III.B of this paper.
\begin{enumerate}
\item
Due to $\mathbf{UU}^H=\mathbf{I}$, Eq. (\ref{samd}) can be achieved via right multiplying the signal space in (\ref{signalspace}) by $\mathbf{U}^H$. Correspondingly, $\mathbf{S}$ is also $K$-row sparse. Since the B5G cellular IoT with massive access is usually interference limited, $\mathbf{V}$ can be exactly approximated as $\mathbf{V}\approx \mathbf{AX}\mathbf{U}^H$ by neglecting the noise term. Because of $r^e=\text{rank}(\mathbf{V})\leq \text{rank}(\mathbf{XU}^H)\leq\text{rank}(\mathbf{X})=r^e$, it is able to obtain that $\text{rank}(\mathbf{S})=\text{rank}(\mathbf{XU}^H)=r^e$. Therefore, $\mathbf{S}$ in the problem (\ref{eq3}) is full column rank and block sparse. It has been proved that if the pilot matrix $\mathbf{A}$ satisfies the block restricted isometry property (Block-RIP), the block-sparse problem (\ref{eq3}) has the bounded estimation error and can stably reconstruct all $K$ row-sparse matrices \cite{block,block1}. For an i.i.d. Gaussian distributed matrix $\mathbf{A}$, it is sufficient to guarantee that $\mathbf{A}$ in (\ref{eq3}) satisfies the Block-RIP under the condition of $L\geq \mathcal{O}(K\ln(N/K)+Kr^e)$. Equivalently, if the length $L$ of pilot sequences meets this lower bound, the $K$-row sparse $\mathbf{S}$ can be stably reconstructed from $\mathbf{V}$.

\item
According to Eq. (\ref{samd}), we have $\mathbf{VU}=\mathbf{A}\mathbf{S}\mathbf{U}+\mathbf{E}_{S}\mathbf{U}$. In other words, the row support of the resulting matrix $\mathbf{S}\mathbf{U}$ corresponds to the row support of ${\mathbf{X}}$. Thus, we can obtain ${\mathbf{X}}$ by letting ${\mathbf{X}}=\mathbf{S}\mathbf{U}$.
\end{enumerate}
\end{IEEEproof}

\emph{Remark 1}: $\mathbf{V}$ preserves the signal space information of $\mathbf{Y}$, hence the device state matrix can be recovered completely. When the BS is equipped with large antenna array in B5G cellular IoT, the reduction of computational complexity is substantial compared with the original one.
Moreover, the dimension reduction also can decrease the required length of pilot sequences. Specifically, the lower bound on the pilot length of unstructured sparse recovery by $l_1$ minimization in \cite{l1norm} is $L\geq \mathcal{O}(KM\ln(N/K))$, and the lower bound is $L\geq \mathcal{O}(K\ln(N/K)+KM)$ for joint-sparse recovery by $l_{21}$ minimization in \cite{block}. The bound of the dimension reduced method is lower than the above ones in practice scenarios.

\subsection{Rank Estimation}
As mentioned above, to perform dimension reduction, it is necessary to obtain the rank of the device state matrix $\mathbf{X}$. In this subsection, we design a rank estimation method according to the characteristic of the received signal $\mathbf{Y}$. First, we provide the following proposition which is instrumental in calculating the essential rank.

\textbf{\bf{Proposition} 1}: Let $\Sigma_{K}$ denote the set of all matrices in $\mathbb{C}^{N\times M} $ having at most $K$'s non-zero rows. If the pilot matrix $\mathbf{A}$ in the mapping $\mathbf{X}\rightarrow \mathbf{A}\mathbf{X} $ is injective on $\Sigma_{K}$, then for every $\mathbf{X}\in \Sigma_{K}$, we have rank $(\mathbf{X})$ = rank $(\mathbf{AX})$.

\begin{IEEEproof}
Please refer to Appendix A.
\end{IEEEproof}

This paper considers the i.i.d. complex Gaussian pilot matrix $\mathbf{A}$ which is injective on $\Sigma_{K}$, thus, according to Proposition 1, the rank of the device state matrix $\mathbf{X}$ is equal to that of the matrix $\mathbf{AX}$.
Adopting the same SVD decomposition method for $\mathbf{Y}$ as in Theorem 1, and considering $\mathbf{AX}$ as the target variable, we can get the decomposition $\mathbf{AX}=\Theta\Xi $ with full column rank matrix $\Theta\in \mathbb{C}^{L\times r^e}$ and matrix $\Xi \in \mathbb{C}^{r^e\times M}$ with $\Xi \Xi ^H=\mathbf{I}$. Therefore, the problem (\ref{eqinter5}) can be rewritten as a general form $\mathbf { {Y}} =  \Theta\Xi +\mathbf {E}$. Since the noise is zero mean and independent of the signal, the covariance matrix of $\mathbf{Y}$ can be expressed as
\begin{equation}\label{COV}
  \mathbf{C}=\mathbb{E}[\Theta\Theta^{H}]+\sigma ^2\mathbf{I}.
\end{equation}
Let $\lambda _i$ denote the $i$th eigenvalue of $\mathbf{C} \in \mathbb{C}^{L\times L}$. Since $\Theta$ is full column rank, we have $\text{rank}\left(\mathbb{E}[\Theta\Theta^{H}]\right)=r^e$, and thus the following relationship holds true: $\lambda _1> \lambda_2> \cdots >\lambda_{r^e}> \lambda_{r^e+1}=\cdots=\lambda_L=\sigma^2$.

Spectral representation theorem states that in the continuous-time case, if $\epsilon (t)$ is a wide-sense stationary
process, then for each fixed $t$, the integral $\epsilon(t)=\int _{-\infty }^{+\infty }\exp(\imath tx)d\zeta(x)$ is the limit in quadratic mean of a
sequence of processes $Q(n,t)=\sum _{h=1}^n\exp(\imath tx_{h-1})[\zeta(x_h)-\zeta(x_{h-1})]$ with some limit integer parameter $n$, a right continuous
and orthogonal-increment process $\zeta(x)$ \cite{spec} and an imaginary unit $\imath $. Since
$\mathbb{E}[\Theta\Theta^{H}]$ is a wide-sense stationary process and it is assumed that the rank of $\mathbb{E}[\Theta\Theta^{H}]$ is the variable $r$, then we can compute a family of covariance matrices model based on Eq. (\ref{COV}) using the above spectral representation theorem as follows:
\begin{equation}
\label{rcov}
\mathbf{C}^{(r)}=\sum_{i=1}^{r}(\lambda_i-\sigma ^2)\mathbf{d}_i\mathbf{d}_i^H+\sigma ^2\mathbf{I},
\end{equation}
where $\lambda_1,\cdots,\lambda_r$ and $\mathbf{d}_1,\cdots,\mathbf{d}_r$ denote the eigenvalues and eigenvectors of the $\mathbf{C}^{(r)}$ at rank $r$ respectively. Here, $r\in \{0, 1,\cdots,L-1\}$ ranges over the set of all possible rank value. Define $\mathbb{Q}^{(r)}=(\lambda_1,\cdots,\lambda_r,\sigma^2,\mathbf{d}_1,\cdots,\mathbf{d}_r)$ as the parameter vector of these models of covariance matrices, and assume that the continuous random vector $\mathbf{y}_m\in \mathbb{C}^{L\times M}$ ($\mathbf{y}_m$ denotes the $m$-th column of $\mathbf{Y}$) obeys the elliptical distribution, which can provide more flexibility in modeling complex received data at the BS with various environment due to the fact that the elliptical family can model the Gaussian data or heavy-tailed data \cite{ellp,ellp1}. Since the received data at different antennas are regarded as statistically independent elliptical distribution, their joint probability density $g(\mathbf{y}_1,\mathbf{y}_2,\cdots,\mathbf{y}_M|\mathbb{Q}^{(r)})$ of the received signal is given by
\begin{equation}\label{CES}
g(\mathbf{y}_1,\mathbf{y}_2,\cdots,\mathbf{y}_M|\mathbb{Q}^{(r)})=\prod _{m=1}^{M}\tau_m\left | \mathbf{C}^{(r)}  \right |^{-1}q(\mathbf{y}_m^H(\mathbf{C}^{(r)})^{-1}\mathbf{y}_m),
\end{equation}
where $q(x)$ is the function of density generator, $|\cdot|$  denotes the determinant of a square matrix, and $\tau_m$ is a normalizing parameter so as to ensure that $\tau_m\left | \mathbf{C}^{(r)}  \right |^{-1}q(\mathbf{y}_m^H(\mathbf{C}^{(r)})^{-1}\mathbf{y}_m)$ integrates to one \cite{ces}. Taking the negative logarithm of probability density results in the negative log-likelihood function, which is given by
\begin{equation}\label{lol}
\mathcal{L}(\mathbb{Q}^{(r)})=\frac{1}{M}\left(\sum_{i=1}^{M}\rho(\mathbf{y}_i^H(\mathbf{C}^{(r)})^{-1}\mathbf{y}_i) -M\ln \left |\mathbf{C}^{(r)}  \right |^{-1}\right),
\end{equation}
where $\rho(x)=-\ln q(x)$. As for convenience later on, we divide the two terms in the right side of (\ref{lol}) by $M$, which does not change the rank estimation. In order to stabilize the rank estimation problem whenever the number of antennas $M$ and the length of pilot $L$ are of similar order in the JADCE problem, a diagonal loading term is introduced into Eq. (\ref{lol}) as follows:
\begin{eqnarray}\label{pen}
\mathcal{L}(\mathbb{Q}^{(r)})&=&\frac{1}{M}\sum_{i=1}^{M}\rho(\mathbf{y}_i^H(\mathbf{C}^{(r)})^{-1}\mathbf{y}_i) -\ln \left |\mathbf{C}^{(r)}  \right |^{-1}\nonumber\\
&&+\beta \text{Tr}((\mathbf{C}^{(r)})^{-1}),
\end{eqnarray}
where $\beta \in(0,1]$ is a regularization parameter, $\text{Tr}$ represents the trace of the matrix. The function $\rho(x)$ can be defined as a general form, not necessarily related to elliptical density distribution $q(x)$ and it is usually chosen such that the derivative of $\rho(x)$ is non-decreasing, non-negative and continuous. In this paper, we consider Rayleigh fading channel and Gaussian distributed noise, thus we set $\rho(x)=(1-\beta)x$. Note that one can flexibly design function $\rho(x)$ for obtaining a good rank estimation according to characteristics of the received data at the BS. Using the property $\text{Tr}(\mathbf{AB})=\text{Tr}(\mathbf{BA})$ and omitting the terms that do not depend on the parameter vector $\mathbf{C}^{(r)}$, Eq. (\ref{pen}) reduces to
\begin{eqnarray}\label{LAS}
  \mathcal{L}(\mathbb{Q}^{(r)})&=&\text{Tr}\left \{  ((1-\beta) \frac{1}{M}\mathbf{Y}\mathbf{Y}^{H}+\beta\mathbf{I})(\mathbf{C}^{(r)})^{-1}\right \} \nonumber \\
  &&-  \ln \left |\mathbf{C}^{(r)}  \right |^{-1}.
\end{eqnarray}
Then we can obtain the estimation $\widehat{\mathbb{Q}}^{(r)}$ of $\mathbb{Q}^{(r)}$ by minimizing Eq. (\ref{LAS}), that is
\begin{eqnarray}
\hat{\lambda}_i&=&\overline{\lambda}_i,~~i=1,\cdots,r\label{EMI} \\
\hat{\mathbf{d}}_i&=&\overline{\mathbf{d}}_i,~~i=1,\cdots,r \label{EMI1}\\
\hat{\sigma}^2&=&\frac{1}{L-r}\sum_{i=r+1}^{L}\overline{\lambda}_i,\label{EMI2}
\end{eqnarray}
where $\overline{\lambda}_1>\overline{\lambda}_2>\cdots>\overline{\lambda}_L$ and the corresponding $\overline{\mathbf{d}}_1, \cdots, \overline{\mathbf{d}}_L$ denote the eigenvalues and eigenvectors of the following matrix
\begin{equation}\label{unmi}
  \hat{\mathbf{C}}=(1-\beta) \frac{1}{M}\mathbf{Y}\mathbf{Y}^{H}+\beta \mathbf{I}.
\end{equation}
Note that Eq. (\ref{unmi}) is a general linear combination estimator of covariance matrices $\mathbf{C}$ elaborated in \cite{bet}. By combining the estimation parameter in Eqs. (\ref{EMI}) - (\ref{EMI2}), we first obtain that

\begin{equation}\label{deter}
\left |\mathbf{C}^{(r)}\right |=\left(\frac{1}{L-r}\sum_{i=r+1}^{L}\overline{\lambda}_i\right)^{(L-r)}\prod _{i=1}^{r}\overline{\lambda}_i,
\end{equation}
and
\begin{eqnarray}\label{deter1}
&&\text{Tr}\left \{  ((1-\beta) \frac{1}{M}\mathbf{Y}\mathbf{Y}^{H}+\beta\mathbf{I})(\mathbf{C}^{(r)})^{-1}\right \} \nonumber \\
&&=(\hat{\sigma}^2)^{-1}(\overline{\lambda}_{r+1}+\overline{\lambda}_{r+2}+\cdots+\overline{\lambda}_L)+r \nonumber\\
&&=L-r+r=L,
\end{eqnarray}
where we use the fact that the product of the eigenvalues of the matrix is equal to its determinant, and the sum of the eigenvalues of the matrix is equal to its trace.
Then, substituting equations (\ref{deter}) and (\ref{deter1}) into (\ref{LAS}), the negative log-likelihood of the parameters estimation $\widehat{\mathbb{Q}}^{(r)}$ can be cast as
\begin{equation}\label{qha}
\mathcal{L}(\widehat{\mathbb{Q}})=L+(L-r)\ln\left(\frac{\sum_{i=r+1}^{L}\overline{\lambda}_i}{L-r}\right)+\sum _{i=1}^r\ln\overline{\lambda}_i,
\end{equation}
where $L$ can be omitted because it is independent of the unknown parameter. Moreover, a penalty term for ensuring an unbiased estimate of the mean Kulback-Liebler distance between $g(\mathbf{y}_1,\mathbf{y}_2,\cdots,\mathbf{y}_M|\mathbb{Q}^{(r)})$ and $g(\mathbf{y}_1,\mathbf{y}_2,\cdots,\mathbf{y}_M|\widehat{\mathbb{Q}}^{(r)})$ is needed. Here, we adopt a penalty $\frac{ur}{M}(L-\frac{r-1}{2})$ controlled by the unspecified constant $u$ and for a detailed account on this penalty, we refer the reader to \cite{nonparam}.
Thus, the rank of $\mathbf{X}$ can be estimated by maximizing the following $CM(r)$ \begin{eqnarray}\label{esre0}
  \hat{r}&=&\mathop \text{argmax}\limits_{r\in\{1,2,\cdots,L\}}CM(r),
\end{eqnarray}
where
\begin{eqnarray}\label{esre}
  CM(r)&=&-(L-r)\ln\left(\frac{\sum_{i=r+1}^{L}\overline{\lambda}_i}{L-r}\right)-\sum _{i=1}^r\ln\overline{\lambda}_i \nonumber \\
  &&-\frac{ur}{M}\left(L-\frac{r-1}{2}\right),
\end{eqnarray}
which is equivalent to selecting the model $\mathbb{Q}^{(r)}$ that best fits the received data $\mathbf{Y}$. The following lemma provides the conditions that guarantee $CM(r)$ is maximized for $r=r^e$.

{\bf{Lemma 1}}: Assume $L, M\rightarrow \infty$ with $L/M \rightarrow \varrho$, and ${\lambda}_1$ is bounded. For $r<r^e$,
if $\lambda_{r^e}+\frac{\varrho\lambda_{r^e}}{\lambda_{r^e}-1}-\sigma^2-\ln(\lambda_{r^e}+\frac{\varrho\lambda_{r^e}}{\lambda_{r^e}-1})>u\varrho$ and $\lambda_{r^e}>1+\sqrt{\varrho}$ hold,
$P_r(CM(r^e)>CM(r))\rightarrow 1$, namely the probability that $CM(r^e)$ is greater than $CM(r)$ tends to 1. For $r>r^e$, if $u>\frac{1-\sigma^2}{\varrho}+1+2\sqrt{1/\varrho}-2\ln(1+\sqrt{\varrho})/\varrho$ holds, $P(CM(r^e)>CM(r))\rightarrow 1$.
\begin{IEEEproof}
Please refer to Appendix B.
\end{IEEEproof}

From Lemma 1, we can observe that accurate rank estimation probability is affected by the transmit power $\lambda_{r^e}-\sigma^2$, and the higher the transmit power, the higher the accurate rank estimation probability. Since $r$ in (\ref{esre0}) is an integer from 1 to $L$, $\hat{r}$ can be obtained by searching the maximum $CM(r)$. The computational complexity of the rank estimation algorithm is $\mathcal{O}(L^2)$, which is low and acceptable. In order to balance the channel estimation error and the efficiency of information transmission, the value of $L$ is generally near the number of active devices in practice, which is not so large in B5G cellular IoT.

\subsection{Joint Activity Detection and Channel Estimation Framework}
In this section, we exploit the characteristic of full column rank of the transformed device state matrix $\mathbf{S}$ for jointly activity detection and channel estimation.

Based on the dimension reduction in Theorem 1, the full rank
information of $\mathbf{S}$ can be incorporated to efficiently determine the solution. Apart from this, we can further utilize the eigenvectors $[\bar{\mathbf{d}}_1,\cdots,\bar{\mathbf{d}}_N]$ of the covariance matrix estimator obtained in the rank estimation to encourage a good activity detection and channel estimation in the noisy environment. Thus, the JADCE problem (\ref{mmv}) can be reformulated as
\begin{eqnarray}\label{OP1}
  \!\!\!\!&&\!\!\!\!\mathop \text{argmin}\limits_{\mathbf{S}}\underbrace{\sum_{n=1}^{N}[\left \| \left(\mathbf{A}^H\overline{\mathbf{D}}\right)[n] \right \|_2*\left \| \mathbf{S}[n] \right \|_{2}]}_{{G}(\mathbf{S})}+\frac{\zeta}{2} \left \| \mathbf{AS}-\mathbf{V} \right \|_F^2\nonumber\\
 \!\!\!\!&&\!\!\!\! \textrm{s.t.}~~~ \mathbf{S}\in \mathbb{C}^{N\times r^e}:\text{rank}(\mathbf{S})=r^e
\end{eqnarray}
where $\left(\mathbf{A}^H\overline{\mathbf{D}}\right)[n]$ is a weighted coefficient assigned to the $n$th device. Note that the vectors $\bar{\mathbf{d}}_{i}$ in Eq. (\ref{EMI1}) can be divided into two components, namely $\widetilde{\mathbf{D}}=\{\bar{\mathbf{d}}_1,\bar{\mathbf{d}}_2,\cdots,\bar{\mathbf{d}}_{r^e} \}$ and $\overline{\mathbf{D}}=[\bar{\mathbf{d}}_{r^e+1},\bar{\mathbf{d}}_{r^e+2},\cdots,\bar{\mathbf{d}}_{L} ]$ corresponding to the signal and its orthogonal spaces, respectively. In the same way, the pilot matrix $\mathbf{A}$ can be divided into $\mathbf{A}_{ac}\in \mathbb{C}^{L\times K}$ and $\mathbf{A}_{na}\in \mathbb{C}^{L\times (N-K)}$ corresponding to active and non-active sets. Define $\mathbf{d}_{i}^*$ as the eigenvectors of real covariance matrix $\mathbf{C}$ and $\mathbf{D}=\{\mathbf{d}_{r^e+1}^*,\mathbf{d}_{r^e+2}^*,\cdots,\mathbf{d}_{L}^* \}$, literature \cite{noise1} has proved that $\mathbf{A}_{ac}^H{\mathbf{D}}= \mathbf{0}$. Thus, when $M\rightarrow \infty$, the following relation
\begin{equation}\label{weig}
  \mathbf{A}^H\overline{\mathbf{D}}=\left[ \begin{array}{l}{\mathbf{A}_{ac}^H\overline{\mathbf{D}}}\\\mathbf{A}_{na}^H\overline{\mathbf{D}}
\end{array} \right]\rightarrow\left[ \begin{array}{l}{\mathbf{0}}\\\mathbf{A}_{na}^H\overline{\mathbf{D}}
\end{array} \right],
\end{equation}
holds true. Even with a limited number of BS antennas, the entries of $\mathbf{A}_{na}^H\overline{\mathbf{D}}$ are usually much larger than $\mathbf{A}_{ac}^H\overline{\mathbf{D}}$ at a wide range of SNR. Therefore, in the problem (\ref{OP1}), we assign small weights to the entries associated with devices that are more likely to be active for improving the detection and estimation accuracy.

Unfortunately, the problem (\ref{OP1}) is non-convex due to the fixed-rank constraint, for which the conventional convex relaxation approaches are inapplicable. To tackle the challenge, we apply the Riemannian optimization method which projects the optimization problem with a constraint $\text{rank}(\mathbf{S})=r^e$ in the Euclidean space onto that in a manifold space. In general, Riemannian optimization requires that the objective function is smooth, but the objective function in problem (\ref{OP1}) is nonsmooth due to the weighted $l_{21}$ norm term of $G(\mathbf{S})$. In the following, we propose a logarithmic smoothing method to smooth the objective function. Specifically, we replace $\left\| \mathbf{S}[n]\right\|_{2}$ with $J(\mathbf{S}[n])$, which is defined as
\begin{equation}\label{logr}
  J( \mathbf{x})=\left\| \mathbf{x}\right\|_{2}-\frac{1}{\theta }\ln\left(1+\theta\left\| \mathbf{x}\right\|_{2}\right),
\end{equation}
where $\theta>0$ is a tunable parameter. This function can solve the nonsmooth problem. Please refer to Appendix C for the detail proof.

Calculating the Maclaurin series of $\ln\left(1+\theta\left\| \mathbf{x}\right\|_{2}\right)$, for a small $\left\| \mathbf{x}\right\|_{2}$, i.e. $\theta\left\| \mathbf{x}\right\|_{2}\leq 1$, we seek to use the second-order statistics of the $\left \| \mathbf{x} \right \|_{2}$ and other terms are negligible for their small value. Therefore, smoother (\ref{logr}) approaches to
\begin{eqnarray}\label{mak}
\!\!\!\!\!\!&&\!\!\!\!\!\!J( \mathbf{x})=\left\| \mathbf{x}\right\|_{2}-\frac{1}{\theta }\left(\theta \left\| \mathbf{x}\right\|_{2}-\frac{\theta^2}{2}\left\| \mathbf{x}\right\|_{2}^2 \right.\nonumber \\
\!\!\!\!\!\!&&\!\!\!\!\!\!\left.+\frac{\theta^3}{3}\left\| \mathbf{x}\right\|_{2}^3-\cdots \right)
\rightarrow \frac{\theta}{2}\left\| \mathbf{x}\right\|_{2}^2, ~\text{as}~\left\| \mathbf{x}\right\|_{2}\rightarrow 0
\end{eqnarray}
while for a relatively large $\left\| \mathbf{x}\right\|_{2}$, the introduced smoothing operator intrinsically uses its lower-order statistics, due to the decreasing weight of the logarithmic term with the increased variable amount. Thus, smoother (\ref{logr}) approximates to
\begin{eqnarray}\label{lar}
  J_\theta(\mathbf{x})\!\!\!\!\!\!&&\!\!\!\!\!\!=\left\| \mathbf{x}\right\|_{2}-\frac{1}{\theta }\ln\left(1+\theta\left\| \mathbf{x}\right\|_{2}\right)\nonumber \\
  \!\!\!\!\!\!&&\!\!\!\!\!\!\rightarrow \left\| \mathbf{x}\right\|_{2},~\text{as}~\left\|\mathbf{x}\right\|_{2}~\rightarrow \infty.
\end{eqnarray}

\begin{figure}[htbp]
\centering
\subfigure[Logarithm smoother.]{
\begin{minipage}[t]{4.1cm}
\centering
\includegraphics[width=2in]{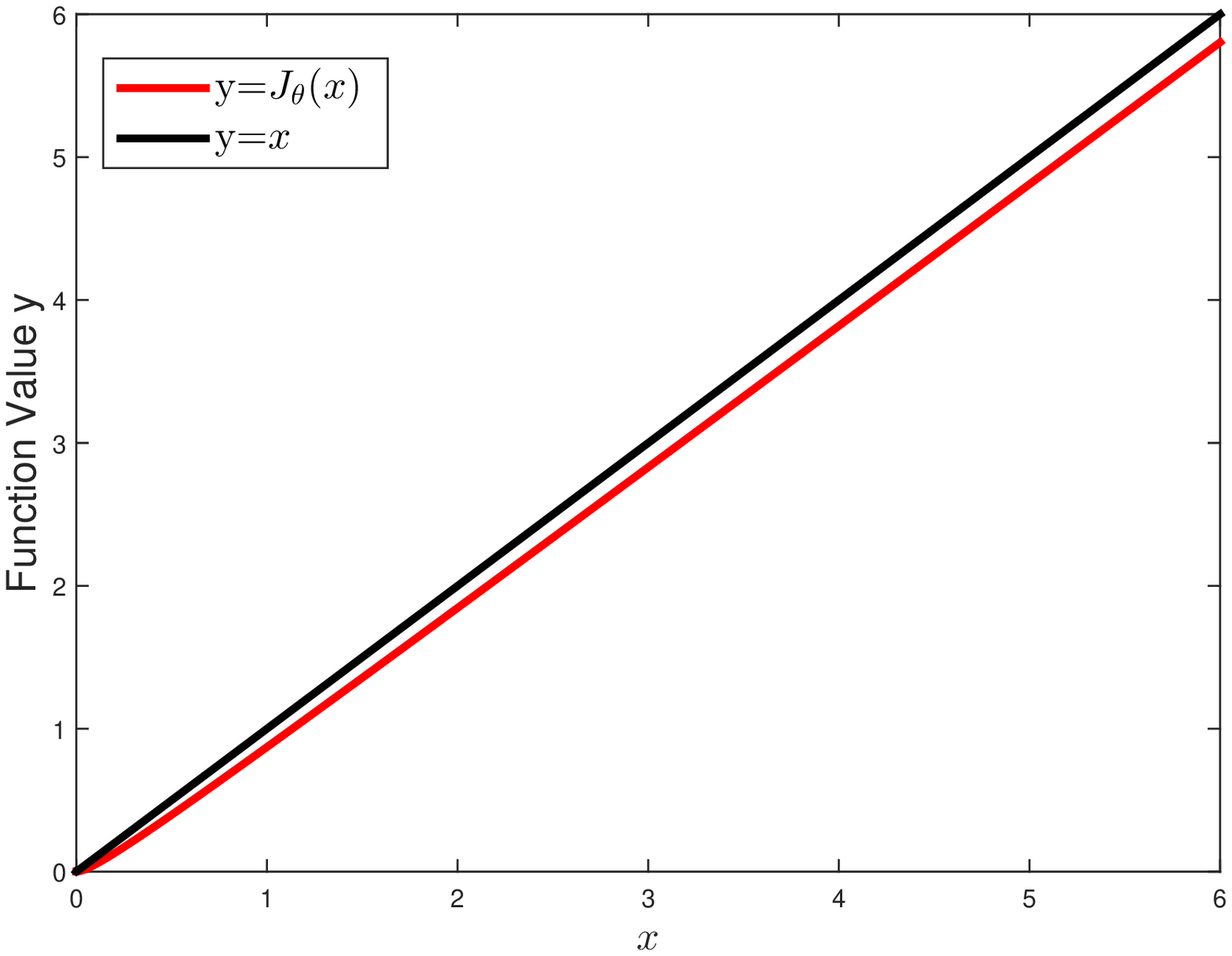}
\end{minipage}%
}%
\subfigure[Derivative of logarithm smoother.]{
\begin{minipage}[t]{6cm}
\centering
\includegraphics[width=2in]{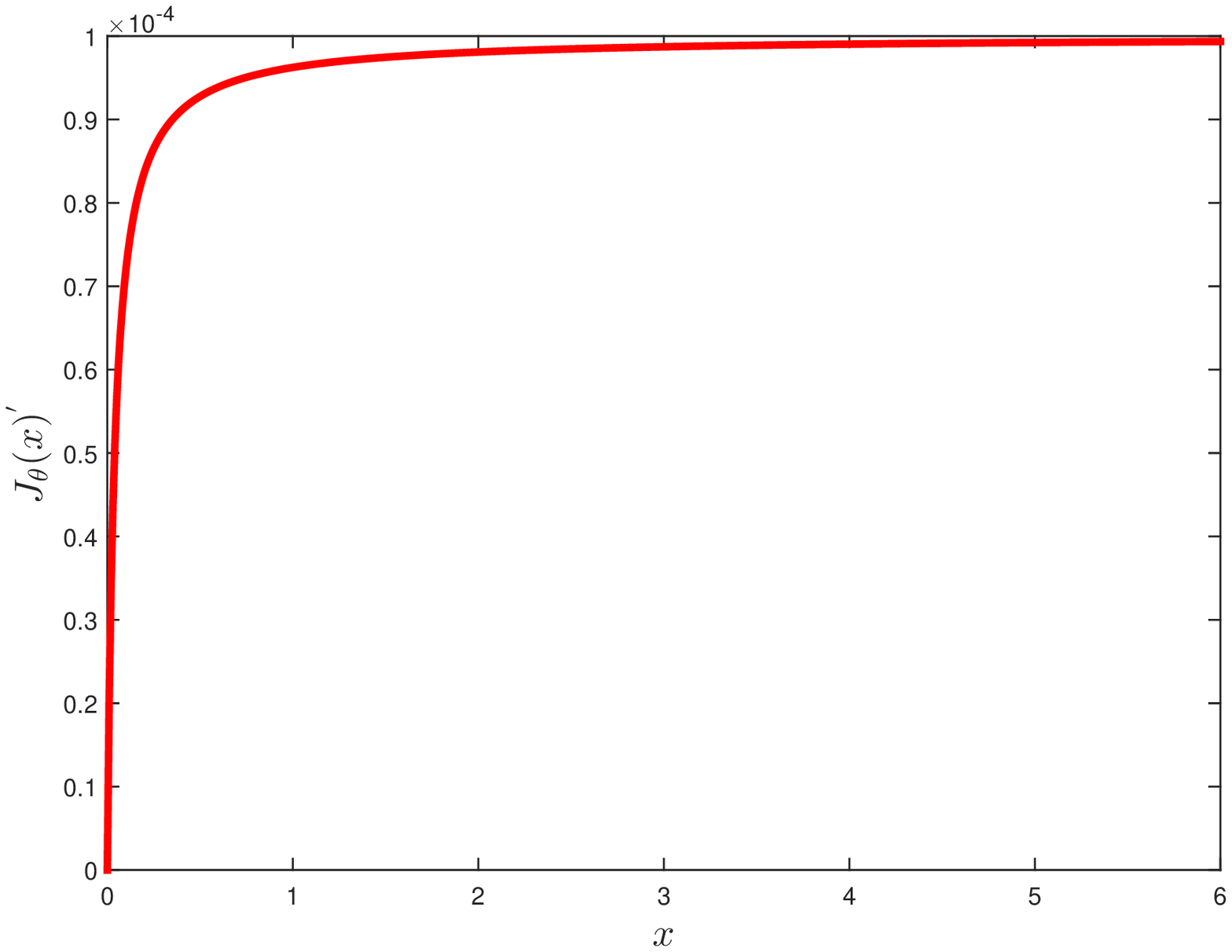}
\end{minipage}%
}%
\centering
\caption{A visualization of smoothing method for $\theta=1/0.039$.}
\label{loga}
\end{figure}

Fig. \ref{loga} shows a visualization of the smoothing method, where $x$ denotes the variable $ \left\| \mathbf{x}\right\|_{2} $. By using such a relative smooth measure, we adjust
the original ${G}(\mathbf{S})$ in problem (\ref{OP1}) elegantly and gradually based on the $\left \| \mathbf{S}[n]\right \|_{2}$ amount. Thus, the problem (\ref{OP1}) can be transformed as

\begin{eqnarray}\label{OP2}
  \!\!\!\!\!\!&&\!\!\!\!\!\!\mathop \text{argmin}\limits_{\mathbf{S}} \underbrace{\sum_{n=1}^{N}\left[\left \| \left(\mathbf{A}^H\overline{\mathbf{D}}\right)[n] \right \|_2*J\left(\left \| \mathbf{S}[n] \right \|_{2}\right)\right]}_{{G}_{\theta}(\mathbf{S})}\nonumber\\
  \!\!\!\!\!\!&&\!\!\!\!\!\!~~~~~~~~~~~~~~~~~~+\frac{\zeta}{2} \left \| \mathbf{AS}-\mathbf{V} \right \|_F^2\nonumber\\
   \!\!\!\!\!\!&&\!\!\!\!\!\! \textrm{s.t.}~~~ \mathbf{S}\in \mathbb{C}^{N\times r^e}:\text{rank}(\mathbf{S})=r^e.
\end{eqnarray}
Note that ${G}_{\theta}(\mathbf{S})$ is an efficient smoothing function of ${G}(\mathbf{S})$ in the whole range and the following theorem states that problem (\ref{OP1}) can be well fitted  by problem (\ref{OP2}).

\textbf{Theorem 2}: Suppose that the problem (\ref{OP1}) has a unique $K$-sparse solution. Then, the problem (\ref{OP2}) satisfying $\theta>\theta_0$ has a unique $K$-sparse solution, where $\theta_0$ is a constant, and the solution of problem (\ref{OP2}) will converge to that of the problem (\ref{OP1}) as $\theta\rightarrow\infty$.

\begin{IEEEproof}
${G}_{\theta}(\mathbf{S})$ and ${G}(\mathbf{S})$ are continuous functions on feasible region. At the same time, ${G}_{\theta}(\mathbf{S})$ is an increasing sequence with respect to $\theta$ and it converges pointwise to a function ${G}(\mathbf{S})$, then according to Proposition 5.4 and Remark 5.5 in \cite{tao}, we have ${G}_{\theta}({\mathbf{S}})$ $\Gamma $-converges to $G(\mathbf{S})$.
\end{IEEEproof}

\emph{Remark 2}: The proposed logarithm smoothing method intrinsically combines the functions with different order of powers in a continuous manner into a single update, which avoids possible anomalies that may arise due to the breaking point in the smooth term with piecewise-function, such as Huber function-based smoothing method introduced in \cite{huber}.

The problem (\ref{OP2}) can be solved by a Riemannian optimization method, which will be discussed in Section IV in detail. Once obtaining the solution $\hat{\mathbf{S}}$ of the problem (\ref{OP2}), the original device state matrix can be recovered by ${\hat{\mathbf{X}}}=\hat{\mathbf{S}}\mathbf{U}$. Afterward, we can detect the device activity by defining the following activity detector.

\emph{Definition 1}: Based on $\hat{\mathbf{X}}$, we define the activity detector as follows
\begin{eqnarray}\label{thr}
k=\begin{cases}
1, & \text{ if } \left \| \hat{\mathbf{X}}(k,:) \right \|^2\geq v^2M \\
0, & \text{ if } \left \| \hat{\mathbf{X}}(k,:)\right \|^2< v^2M
\end{cases}
\end{eqnarray}
with $v=v_1\max(\hat{\mathbf{X}}(n,m)), \forall n\in N,m \in M$, where $\max(\hat{\mathbf{X}}(n,m))$ is the operation extracting the maximum element value from $\hat{\mathbf{X}}$, $v_1=0.1$ denotes the ratio of the minimum and maximum amplitudes of the channel coefficients.

Meanwhile, the CSI corresponding to active devices can be estimated as
\begin{equation}
\label{gainesti}
\hat{\mathbf{h}}_{k}=\hat{\mathbf{x}}_{k}/\sqrt{\varsigma_k}, \forall k\in\hat{\mathcal{K}}.
\end{equation}
In summary, the proposed dimension reduction-based JADCE algorithm (DR-JADCE) can be described as Algorithm 1.

\begin{algorithm}[h]
\caption{Dimension Reduction-Based Joint Device Detection and Channel Estimation via Riemnnian Optimization}
\label{alg1}
\begin{algorithmic}[1]
\STATE \textbf{Input}: The pilot matrix $\mathbf{A}$, the signal measurements $\mathbf{Y}$, transmit energy $\varsigma_n, \forall n \in \{1,2,\dots,N \}$.
\STATE \textbf{Dimension reduction}: $\hat{r}=\mathop \text{argmin}\limits_{r}{CM}(r)$, $\mathbf{Y}\rightarrow\mathbf {VU}$\STATE \textbf{Device detection}: Solve (\ref{OP2}) with the Algorithm 3 in Section IV to obtain the estimation $\hat{\mathbf{S}}$ of $\mathbf{S}$;
~~~~~~~~~~~~~~~~~~~~~~\text{Recovery the original interested signal}: ${\hat{\mathbf{X}}}=\hat{\mathbf{S}}\mathbf{U}$;\\
Threshold elements of $\hat{\mathbf{X}}$: $\hat{\mathcal{K}}=\left \{ k: \left \| \hat{\mathbf{X}}(k,:)\right \|^2\geq v^2 M\right \}$.
\STATE \textbf{Channel estimation}:
Return $\hat{\mathbf{h}}_{k}=\hat{\mathbf{x}}_{k}/\sqrt{\varsigma_k},~\forall k \in \hat{\mathcal{K}}$
\STATE \textbf{Output}: The estimated support of active devices $\hat{\mathcal{K}}$ and the estimated channel vector ${\hat{\mathbf{h}}}_{\hat{\mathcal{K}}}$.
\end{algorithmic}
\end{algorithm}

\emph{Remark 3}: Intuitively, rank estimation error may occur, which affects the accuracy of activity detection and channel estimation. Fortunately, as will be verified by simulations in Section V, activity detection error and channel estimation error of the proposed algorithm are not sensitive to the rank estimation accuracy when underestimating the rank in the short pilot region. Inspired by these observations, even if the actual rank is known, we can utilize a small rank in the proposed algorithm to further reduce the computational complexity.

\section{Riemannian Optimization for JADCE}
The existing Riemannian algorithms for fixed-rank matrix optimization problems do not work for the problem (\ref{OP2}) with the non-square unknown matrix on the complex manifold. In this section, we first reformulate the resulting rank-constrained smoothing non-convex optimization problem in a specific way, and then develop a Riemannian trust-region algorithm to solve the reformulated problem in the complex field.

\subsection{Problem Reformulation}
To transform the original interested matrix in JADCE to a positive semidefinite matrix for exploiting the specific quotient manifold, we propose to rephrase the problem (\ref{OP2}). Specifically, any rank-$r^e$ positive semidefinite matrix $\mathbf{L}\in \mathbb{C}^{(N+r^e)\times (N+r^e)}$ admits a factorization $\mathbf{L}=\mathbf{Z}\mathbf{Z}^H$ with full column rank matrix $\mathbf{Z}\in \mathbb{C}^{(N+r^e)\times r^e}$.
First, let us define $\mathbf{Z}=\left[ \begin{array}{l}
\mathbf{J} \\
\widetilde{\mathbf{J}}
\end{array} \right]$
with full column-rank matrices $\mathbf{J}\in \mathbb{C}^{N\times r^e}$ and $\widetilde{\mathbf{J}} \in \mathbb{C}^{r^e\times r^e}$. Then utilizing the factorization on $\mathbf{S}$, i.e. $\mathbf{S}=\mathbf{J}\widetilde{\mathbf{J}}^H$ , we can lift $\mathbf{S}$ in a factored form as follows:
\begin{equation}\label{FAC}
  \mathbf{L}=\mathbf{Z}\mathbf{Z}^H=\left[ \begin{array}{l}
\mathbf{J}\mathbf{J}^H ~~~\mathbf{J}\widetilde{\mathbf{J}}^H\nonumber\\
\widetilde{\mathbf{J}}\mathbf{J}^H~~~\widetilde{\mathbf{J}}\widetilde{\mathbf{J}}^H
\end{array} \right].
\end{equation}
In addition to this redefinition, we also introduce two auxiliary matrices $\mathbf{\overline{P}}$ and $\mathbf{\widetilde{P}}$ as
\begin{equation}\label{barp}
\mathbf{\overline{P}}=\left [ \mathbf{I}_N~~~\mathbf{0} \right ]\in\mathbb{C}^{N\times (N+r^e)},
\end{equation}
and
\begin{equation}\label{tidep}
 \mathbf{\widetilde{P}}=\left[ \begin{array}{l}
\mathbf{0} \\
\mathbf{I}_{r^e}
\end{array} \right]\in\mathbb{C}^{ (N+r^e)\times r^e},
\end{equation}
where the blocks $\mathbf{I}_N$ and $\mathbf{I}_{r^e}$ denote the identity matrices of order $N$ and $r^e$, respectively. Upon multiplying both sides of $\mathbf{Z}\mathbf{Z}^H$ by $\mathbf{\overline{P}}$ and $\mathbf{\widetilde{P}}$, we obtain $\mathbf{S}=\mathbf{J}\widetilde{\mathbf{J}}^H=\mathbf{\overline{P}}\mathbf{Z}\mathbf{Z}^H\mathbf{\widetilde{P}}$.
Consequently, the problem (\ref{OP2}) reduces to
\begin{eqnarray}\label{lift}
\!\!\!\!&&\!\!\!\!\mathop \text{argmin}\limits_{\mathbf{Z}}f(\mathbf{Z})=\sum_{n=1}^{N}\left[\left \| \left(\mathbf{A}^H\overline{\mathbf{D}}\right)[n] \right \|_2*J\left(\left \| \left(\mathbf{\overline{P}}\mathbf{Z}\mathbf{Z}^H\mathbf{\widetilde{P}}\right)[n] \right \|_{2}\right)\right]\nonumber \\
\!\!\!\!&&\!\!\!\!~~~~~~~~~~~~~~~~~+\frac{\zeta}{2} \left \| \mathbf{A}\mathbf{\overline{P}}\mathbf{Z}\mathbf{Z}^H\mathbf{\widetilde{P}}-\mathbf{V} \right \|_F^2\nonumber\\
\!\!\!\!&&\!\!\!\!\textrm{s.t.}~~~ \mathbf{Z}\in \mathbb{C}^{(N+r^e)\times r^e}:\text{rank}(\mathbf{Z})=r^e
\end{eqnarray}

Next, we define a non-compact stiefel manifold $\overline{\mathcal{M}}=\left \{ \mathbf{Z}\in \mathbb{C}^{(N+r^e)\times r^e}:\text{rank}(\mathbf{Z})=r^e \right \}$, which denotes the set of all $(N+r^e)\times r^e$ matrices whose columns are linearly independent. Then the rephrased problem (\ref{lift}) can be recast as a Riemannian optimization problem over complex non-compact stiefel manifold. Herein, we first recovery the unknown matrix $\mathbf{Z}$. After the solution $\hat{\mathbf{Z}}$ of problem (\ref{lift}) is obtained, the original solution $\hat{\mathbf{S}}$ can be extracted from $\hat{\mathbf{Z}}$ by the operation $\hat{\mathbf{S}}=\mathbf{\overline{P}}\hat{\mathbf{Z}}(\hat{\mathbf{Z}})^H\mathbf{\widetilde{P}}$.
Although this operation increases the dimension of interested matrix from $\mathbf{S}\in\mathbb{C}^{N\times r^e}$ to $\mathbf{Z}\in\mathbb{C}^{(N+r^e)\times r^e}$, this kind of transform involves a series of sparse matrix multiplication which suggests that the computation complexity is low.  In the following, we shall take advantage of the particular geometric structure of (\ref{lift}) to develop a Riemannian trust-region algorithm.

\subsection{Manifold Geometric Structure of The Parameter $\mathbf{Z}$}
A key property of $\mathbf{L}=\mathbf{Z}\mathbf{Z}^H$ in rephrased problem (\ref{lift}) is that it is invariant over the projection
$\mathbf{Z}\mapsto \mathbf{Z}\mathbf{Q}$, where $\mathbf{Q}\in \mathcal{U}(r^e) =\left \{ \mathbf{Q}\in \mathbb{C}^{r^e\times r^e}:\mathbf{Q}^H\mathbf{Q}=\mathbf{Q}\mathbf{Q}^H=\mathbf{I} \right \}$. This symmetry comes from the invariant relation $\mathbf{Z}\mathbf{Z}^H=\mathbf{Z}\mathbf{Q}(\mathbf{Z}\mathbf{Q})^H$. This means that for a solution $\mathbf{Z}$ to (\ref{lift}), $\mathbf{ZQ}$ is also a
feasible solution. To address this indeterminacy, we consider a set of equivalence classes defined as
\begin{equation}\label{equi}
  [\mathbf{Z}]=\left \{\mathbf{Z}\mathbf{Q}: \mathbf{Q}\in \mathcal{U}(r^e) \right \},
\end{equation}
which encodes the invariance map in an abstract search space called the quotient space, which is denoted as
\begin{equation}\label{quo}
  \mathcal{M}:=\overline{\mathcal{M}}/\mathcal{U}(r^e),
\end{equation}
where the non-compact stiefel manifold $\overline{\mathcal{M}}$ is regarded as the total space.
Consequently, the problem (\ref{lift}) is now reformulated as the following unconstrained optimization problem over the set of equivalence classes in (\ref{equi})
\begin{equation}\label{sm}
  \mathop \text{argmin}\limits_{[\mathbf{Z}]\in \mathcal{M}}f([\mathbf{Z}]).
\end{equation}
Therefore, through optimizing the problem over $\mathcal{M}$, the invariance can be well addressed.

To describe the manifold in Euclidean space, we first linearize the search space by utilizing the tangent space. Specifically, the tangent space to $\overline{\mathcal{M}} $ at $\mathbf{Z}$ is given by $ \mathcal{T}_\mathbf{Z}\overline{\mathcal{M}}$, which is the set of all tangent vectors to $\overline{\mathcal{M}} $ at $\mathbf{Z}$ \cite{pa}. However, due to the fact that the manifold $\mathcal{M}$ is an abstract space, the elements of its tangent space $\mathcal{T}_\mathbf{Z}\mathcal{M}$ need a matrix representation in the total space $\overline{\mathcal{M}}$. Thus $\mathcal{T}_\mathbf{Z}\overline{\mathcal{M}}$ is decomposed into the sum of two complementary spaces such that
\begin{equation}\label{TWO}
  \mathcal{T}_\mathbf{Z}\overline{\mathcal{M}}=\mathcal{V}_{\mathbf{Z}}\oplus \mathcal{H}_{\mathbf{Z}},
\end{equation}
where $\oplus$ is the direct sum of two subspace, the vertical space $\mathcal{V}_{\mathbf{Z}}$ denotes
the directions tangential to the equivalence class $[\mathbf{Z}]$, while the horizontal space $\mathcal{H}_{\mathbf{Z}}$, which is orthogonal to the set of equivalence classes, provides us a valid matrix representation of the abstract tangent space $\mathcal{T}_{\mathbf{Z}}\mathcal{M}$. In this context, for any $\overline{\boldsymbol{\xi}}_{\mathbf{Z}} \in \mathcal{T}_{\mathbf{Z}}{\mathcal{M}}$, there exists a unique horizontal lift $\boldsymbol{\xi}_{\mathbf{Z}}\in \mathcal{H}_{\mathbf{Z}}$ satisfying $
  \boldsymbol{\xi}_{\mathbf{Z}}:=\Pi_{\mathbf{Z}}^h (\overline{\boldsymbol{\xi}}_{\mathbf{Z}})$,
where $\Pi_{\mathbf{Z}}^h $ denotes the projection from $\mathcal{T}_{\mathbf{Z}}{\mathcal{M}}$ onto the horizontal space $\mathcal{H}_{\mathbf{Z}}$ at $\mathbf{Z}$.

\begin{figure}[h]
  \centering
\includegraphics [width=0.35\textwidth] {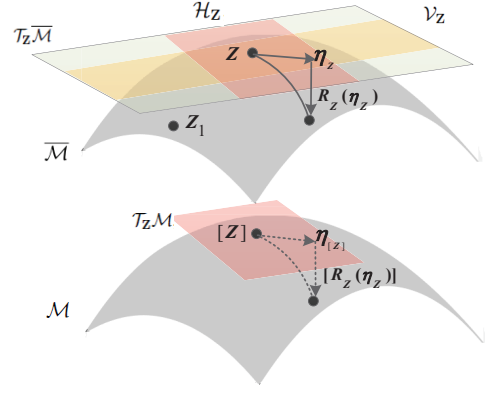}
\caption{Illustrations of the tangent space and Riemannian retraction.}
\label{riemannian}
\end{figure}

Fig. \ref{riemannian} illustrates the tangent space above and Riemannian retraction which will be discussed in the following subsection. The black points $\mathbf{Z}$ and $\mathbf{Z}_1$ on $\overline{\mathcal{M}}$ belong to the equivalence class of solutions relating to $\mathbf{Z}$ by a unitary ambiguity. They are represented by a single point $[\mathbf{Z}]$ on the quotient manifold ${\mathcal{M}}$.

Now the key point is to find the vertical space, horizontal space and the horizontal projection. In detail, we first define the vertical space $\mathcal{V}_{\mathbf{Z}}$ as follows:
\begin{equation}\label{vers}
  \mathcal{V}_{\mathbf{Z}}=\left \{ \mathbf{Z}\mathbf{B}: \mathbf{B}^H=-\mathbf{B}, \mathbf{B}\in\mathbb{C}^{r^e\times r^e}\right \}.
\end{equation}
Correspondingly, the horizontal space can be derived from the following definition
\begin{equation}\label{HOF}
   \mathcal{H}_{\mathbf{Z}}=\left \{ \boldsymbol{\xi}_{\mathbf{Z}}\in \mathcal{T}_\mathbf{Z}\overline{\mathcal{M}}:g(\boldsymbol{\xi}_{\mathbf{Z}},\boldsymbol{\eta}_{\mathbf{Z}})=0,\forall \boldsymbol{\eta}_{\mathbf{Z}}\in \mathcal{V}_{\mathbf{Z}}\right \},
\end{equation}
where $g(\cdot)$ denotes the Riemannian metric for the manifold $\overline{\mathcal{M}}$, which is the smoothly varying inner product: $\mathcal{T}_\mathbf{Z}\overline{\mathcal{M}}\times \mathcal{T}_\mathbf{Z}\overline{\mathcal{M}} \mapsto  \mathbb{R}$. In this paper, we set Riemannian metric as
\begin{eqnarray}\label{inner}
  g_{\mathbf{Z}}(\boldsymbol{\xi}_{\mathbf{Z}},\boldsymbol{\eta}_{\mathbf{Z}})\!\!\!\!\!\!&=&\!\!\!\!\!\!\text{Tr}(\Re (\boldsymbol{\xi}_{\mathbf{Z}}^H\boldsymbol{\eta}_{\mathbf{Z}}) \nonumber\\
\!\!\!\!\!\!&=&\!\!\!\!\!\!\frac{1}{2}\text{Tr}(\boldsymbol{\xi}_{\mathbf{Z}}^H\boldsymbol{\eta}_{\mathbf{Z}}+\boldsymbol{\eta}_{\mathbf{Z}}^H\boldsymbol{\xi}_{\mathbf{Z}}),\boldsymbol{\xi}_{\mathbf{Z}},\boldsymbol{\eta}_{\mathbf{Z}} \in \mathcal{T}_\mathbf{Z}\overline{\mathcal{M}},
\end{eqnarray}
which is equivalent to treating $\mathbb{C}^{N\times r^e}$ as $\mathbb{R}^{2N\times 2r^e}$ with the canonical inner product, and where $\Re(\mathbf{B})$ denotes real part of $\mathbf{B}$. Now, $\mathcal{T}_\mathbf{Z}\overline{\mathcal{M}}$ endowed with inner product leads to a Riemannian manifold $\overline{\mathcal{M}}$. Following the
results above, we can obtain the concise expressions of $\mathcal{H}_{\mathbf{Z}}$ and $\Pi_{\mathbf{Z}}^h$. The horizontal space at $\mathbf{Z}$ is given by
\begin{equation}\label{hois}
  \mathcal{H}_{\mathbf{Z}}=\left \{ \boldsymbol{\xi}_{\mathbf{Z}}\in \mathbb{C}^{(N+r^e)\times r^e}: \boldsymbol{\xi}_{\mathbf{Z}}^H\mathbf{Z}=\mathbf{Z}^H\boldsymbol{\xi}_{\mathbf{Z}}\right \},
\end{equation}
and the projection of any direction $\overline{\boldsymbol{\xi}}_{\mathbf{Z}}$ onto the horizontal space at $\mathbf{Z}$ is given by
\begin{equation}\label{pro1}
  \Pi_{\mathbf{Z}}^h(\overline{\boldsymbol{\xi}}_{\mathbf{Z}})=\overline{\boldsymbol{\xi}}_{\mathbf{Z}}-\mathbf{Z}\mathbf{B},
\end{equation}
where $\mathbf{B}$ is a complex matrix of size $r^e \times r^e$, which is the solution of the following Lyapunov equation
\begin{equation}\label{ome}
\mathbf{Z}^H\mathbf{Z}\mathbf{B}+\mathbf{B}\mathbf{Z}^H\mathbf{Z}=\mathbf{Z}^H\overline{\boldsymbol{\xi}}_{\mathbf{Z}}-\overline{\boldsymbol{\xi}}_{\mathbf{Z}}^H\mathbf{Z}.
\end{equation}
Please refer to Appendix D for the detail proof.

\subsection{Riemannian Gradient and Hessian for JADCE Problem}
We now use the notions developed in the previous section to deduce Riemannian gradient and Hessian on manifolds represented in the tangent space to minimize the cost function $f(\mathbf{Z})$ on the $\mathcal{M}$.

The Riemannian gradient $\overline{\text{grad}f(\mathbf{Z})}$ of $f(\mathbf{Z})$ is the unique operator satisfying
\begin{equation}\label{ggrad}
  g_{\mathbf{Z}}(\overline{\text{grad}f(\mathbf{Z})},\boldsymbol{\xi}_{\mathbf{Z}})=Df (\boldsymbol{Z})[\boldsymbol{\xi}_{\mathbf{Z}}],\forall \boldsymbol{\xi}_{\mathbf{Z}}\in \mathcal{T}_\mathbf{Z}\overline{\mathcal{M}},
\end{equation}
where
 \begin{equation}\label{yt}
Df(\boldsymbol{Z})[\boldsymbol{\xi}_{\mathbf{Z}}]:=\lim_{t\rightarrow 0}(f(\mathbf{Z}+t\boldsymbol{\xi}_{\mathbf{Z}})-f(\mathbf{Z}))/t,
 \end{equation}
is the directional derivative of $f(\mathbf{Z})$ in the direction $\boldsymbol{\xi}_{\mathbf{Z}}$.

Substituting the objective function defined in (\ref{lift}) and the Riemannian metric in (\ref{inner}) into (\ref{ggrad}) yields
\begin{eqnarray}\label{gradp}
\!\!\!\!\!\!\!\!\!&&\!\!\!\!\!\!\!\!\!\overline{\text{grad}f(\mathbf{Z})}=\frac{\zeta}{2}\overline{\mathbf{P}}^H\mathbf{A}^H\left(\mathbf{A}\overline{\mathbf{P}}\mathbf{Z}
\mathbf{Z}^H\widetilde{\mathbf{P}}-\mathbf{V}\right)\widetilde{\mathbf{P}}^H\mathbf{Z}\nonumber\\
\!\!\!\!\!\!\!\!\!&&\!\!\!\!\!\!\!\!\!+\frac{\zeta}{2}\widetilde{\mathbf{P}}\left(\mathbf{A}
\overline{\mathbf{P}}\mathbf{Z}\mathbf{Z}^H\widetilde{\mathbf{P}}-\mathbf{V}\right)^H\mathbf{A}\overline{\mathbf{P}}\mathbf{Z}\nonumber \\
\!\!\!\!\!\!\!\!\!&&\!\!\!\!\!\!\!\!\!+\overline{\mathbf{P}}^H\left[ \frac{\left \| \mathbf{A}^H\overline{\mathbf{D}}[1] \right \|_2 \theta\boldsymbol{\Omega }_1^H}{1+\theta\left \| \boldsymbol{\Omega}_1\right \|_2},\cdots,\frac{\left \| \left(\mathbf{A}^H\overline{\mathbf{D}}\right)[N] \right \|_2 \theta\boldsymbol{\Omega }_N^H}{1+\theta\left \| \boldsymbol{\Omega}_N\right \|_2} \right]^H \widetilde{\mathbf{P}}^H\mathbf{Z}\nonumber\\
\!\!\!\!\!\!\!\!\!&&\!\!\!\!\!\!\!\!\!+\widetilde{\mathbf{P}}\left[ \frac{\left \| \mathbf{A}^H\overline{\mathbf{D}}[1] \right \|_2 \theta\boldsymbol{\Omega }_1^H}{1+\theta\left \| \boldsymbol{\Omega}_1\right \|_2},\cdots,\frac{\left \| \left(\mathbf{A}^H\overline{\mathbf{D}}\right)[N] \right \|_2 \theta\boldsymbol{\Omega }_N^H}{1+\theta\left \| \boldsymbol{\Omega}_N\right \|_2} \right]\overline{\mathbf{P}}\mathbf{Z},\nonumber\\
\end{eqnarray}
where $\boldsymbol{\Omega }_n={\mathbf{P}}_n\overline{\mathbf{P}}\mathbf{Z}\mathbf{Z}^H\widetilde{\mathbf{P}}$, $\mathbf{P}_n\in\mathbb{C}^{1\times N}$ is the row selection matrix whose all elements are zero except the $n$th enetry for $n\in \{1, 2,\cdots,N\}$. Please refer to Appendix E for the detail proof.

Then the Riemannian gradient can be uniquely represented by its horizontal lift in $\overline{\mathcal{M}}$ and the corresponding matrix representation is given by
\begin{equation}\label{liftgrad}
{\text{grad}}f(\mathbf{Z})=\Pi_{\mathbf{Z}}^h\left(\overline{\text{grad}f(\mathbf{Z})}\right).
\end{equation}

In order to use the second-order information of the objective functions, which can escape from saddle points then provide more accurate recovery solution, we need to exploit the Riemannian Hessian of $f(\mathbf{Z})$. The horizontal lift of the Riemannian Hessian along given direction $\boldsymbol{\eta}_{\mathbf{Z}}$ has the following matrix expression
\begin{equation}\label{hisrie}
  \text{Hess}f(\mathbf{Z})[\boldsymbol{\eta}_{\mathbf{Z}}]=\Pi_{\mathbf{Z}}^h
  \left(\lim_{t\rightarrow 0}\left(\overline{\text{grad}f(\mathbf{Z}+t{\boldsymbol{\eta}}_{\mathbf{Z}}})-\overline{\text{grad}f(\mathbf{Z}})\right)/t\right).
\end{equation}
Now, applying this formula to the vector field $\overline{\text{grad}f(\mathbf{Z}})$ leads to
\begin{eqnarray}\label{hes}
\!\!\!\!\!\!\!\!\!&&\!\!\!\!\!\!\!\!\!\text{Hess}f(\mathbf{Z})[\boldsymbol{\eta}_{\mathbf{Z}}]\nonumber\\
\!\!\!\!\!\!\!\!\!&&\!\!\!\!\!\!\!\!\!=\Pi_{\mathbf{Z}}^h\left( \frac{\zeta}{2} \left( \overline{\mathbf{P}}^H\mathbf{A}^H\mathbf{A}\overline{\mathbf{P}}\left(\mathbf{Z}\boldsymbol{\eta}_{\mathbf{Z}}^H+
\boldsymbol{\eta}_{\mathbf{Z}}\mathbf{Z}^H\right)\widetilde{\mathbf{P}}\widetilde{\mathbf{P}}^H\mathbf{Z}-\widetilde{\mathbf{P}}\mathbf{V}^H\mathbf{A}
\overline{\mathbf{P}}\boldsymbol{\eta}_{\mathbf{Z}}\right.\right.\nonumber\\
\!\!\!\!\!\!\!\!\!&&\!\!\!\!\!\!\!\!\!\left.\left.+\widetilde{\mathbf{P}}\left(\mathbf{A}
\overline{\mathbf{P}}\left(\mathbf{Z}\boldsymbol{\eta}_{\mathbf{Z}}^H
+\boldsymbol{\eta}_{\mathbf{Z}}\mathbf{Z}^H\right)\widetilde{\mathbf{P}}\right)^H\mathbf{A}\overline{\mathbf{P}}\mathbf{Z} -\overline{\mathbf{P}}^H\mathbf{A}^H\mathbf{V}\widetilde{\mathbf{P}}^H\boldsymbol{\eta}_{\mathbf{Z}} \right.\right. \nonumber\\
\!\!\!\!\!\!\!\!\!&&\!\!\!\!\!\!\!\!\!\left.\left.+  \widetilde{\mathbf{P}}\left(\mathbf{A}\overline{\mathbf{P}}\mathbf{Z}\mathbf{Z}^H\widetilde{\mathbf{P}}\right)^H\mathbf{A}\overline{\mathbf{P}}\boldsymbol{\eta}_{\mathbf{Z}}
+\overline{\mathbf{P}}^H\mathbf{A}^H\mathbf{A}\overline{\mathbf{P}}\mathbf{Z}\mathbf{Z}^H\widetilde{\mathbf{P}}\widetilde{\mathbf{P}}^H\boldsymbol{\eta}_{\mathbf{Z}}\right)\right.\nonumber \\
\!\!\!\!\!\!\!\!\!&&\!\!\!\!\!\!\!\!\!\left.+\left[ \frac{\left \| \mathbf{A}^H\overline{\mathbf{D}}[1] \right \|_2\theta\mathbf{\widetilde{P}} \boldsymbol{\Omega }_1^H}{1+\theta\left \| \boldsymbol{\Omega }_1 \right \|_2},\cdots , \frac{\left \| \left(\mathbf{A}^H\overline{\mathbf{D}}\right)[N] \right \|_2\theta \mathbf{\widetilde{P}}\boldsymbol{\Omega }_1^H}{1+\theta\left \| \boldsymbol{\Omega }_N \right \|_2}\right]\overline{\mathbf{P}}\boldsymbol{\eta}_{\mathbf{Z}} \right. \nonumber \\
\!\!\!\!\!\!\!\!\!&&\!\!\!\!\!\!\!\!\!\left.+\mathbf{\overline{P}}^H\left[ \frac{\left \| \mathbf{A}^H\overline{\mathbf{D}}[1] \right \|_2 \theta\boldsymbol{\Omega }_1^H}{1+\theta\left \| \boldsymbol{\Omega}_1\right \|_2},\cdots,\frac{\left \| \left(\mathbf{A}^H\overline{\mathbf{D}}\right)[N] \right \|_2 \theta\boldsymbol{\Omega }_N^H}{1+\theta\left \| \boldsymbol{\Omega}_N\right \|_2} \right]^H\widetilde{\mathbf{P}}^H\boldsymbol{\eta}_{\mathbf{Z}}\right.\nonumber\\
\!\!\!\!\!\!\!\!\!&&\!\!\!\!\!\!\!\!\!\left.+\mathbf{\overline{P}}^H\left[ \mathbf{\Gamma}_1^H,\cdots ,\mathbf{\Gamma} _N^H \right]^H \widetilde{\mathbf{P}}^H\mathbf{Z}+\mathbf{\widetilde{P}}\left[ \mathbf{\Gamma}_1^H,\cdots ,\mathbf{\Gamma} _N^H \right]\overline{\mathbf{P}}\mathbf{Z}\right),
\end{eqnarray}
where
\begin{eqnarray}\label{lamm}
\!\!\!\!\!\!&&\!\!\!\!\!\!\mathbf{\Gamma} _n=\frac{1}{\left( 1+\theta\left \| \boldsymbol{\Omega }_n \right \|_2 \right )^2}\left(\left \| \left(\mathbf{A}^H\overline{\mathbf{D}}\right)[n] \right \|_2\theta \mathbf{\Lambda} _n(1+\theta\left\| \boldsymbol{\Omega }_n \right\|_2)\right.\nonumber\\
\!\!\!\!\!\!&&\!\!\!\!\!\!\left.-0.5\theta\left \| \boldsymbol{\Omega }_n \right \|_2^{-1}\left(\boldsymbol{\Omega }_n\mathbf{\Lambda} _n^H+\mathbf{\Lambda}_n\boldsymbol{\Omega }_n^H\right)\left \| \left(\mathbf{A}^H\overline{\mathbf{D}}\right)[n] \right \|_2\theta\boldsymbol{\Omega }_n\right),\nonumber\\
\end{eqnarray}
and
\begin{equation}\label{G}
  \mathbf{\Lambda} _n={\mathbf{P}}_n\overline{\mathbf{P}}(\mathbf{Z}\boldsymbol{\eta}_{\mathbf{Z}}^H
  +\boldsymbol{\eta}_{\mathbf{Z}}\mathbf{Z}^H)\widetilde{\mathbf{P}},
\end{equation}
where $n\in \{1, 2,\cdots,N\}$.

\subsection{Riemannian Trust-Region Optimization for JADCE}
With the Riemannian gradient and Hessian at hand, we now need to determine the search direction in the tangent space $\mathcal{T}_\mathbf{Z}\overline{\mathcal{M}}$ and a retraction that can map the search direction from $\mathcal{H}_{\mathbf{Z}}$ to $\overline{\mathcal{M}}$. In the $t$-th iterate of the problem, the following efficient retraction is adopted for ensuring that each update of our search variable is located on the manifold:
\begin{equation}\label{sta}
  \mathbf{Z}_{t+1}=\mathbf{R}_{\mathbf{Z}}(\boldsymbol{\eta}_{\mathbf{Z}}^t)
    =\mathbf{Z}_{t}+\alpha_t\boldsymbol{\eta}_{\mathbf{Z}}^t,
\end{equation}
where $\alpha_t$ is the step size, $\boldsymbol{\eta}_{\mathbf{Z}}^t\in \mathcal{H}_{\mathbf{Z}}$ is a search direction. Such a retraction can provide a computationally efficient way to smoothly select a moving curve on a manifold. Eq. (\ref{sta}) can be translated into the update $[\mathbf{Z}_{t+1}]=[\mathbf{R}_{\mathbf{Z}}(\boldsymbol{\eta}_{\mathbf{Z}}^t)]$ on $\mathcal{M}$ as illustrated in Fig. {\ref{riemannian}}.

We now gather the Riemannian gradient and Hessian to derive a search direction $\boldsymbol{\eta}_\mathbf{Z}^t$ in problem (\ref{sta}) by solving the following trust-region subproblem
\begin{eqnarray}\label{trust}
  \!\!\!\!&&\!\!\!\!\mathop \text{argmin}\limits_{\boldsymbol{\eta}_\mathbf{Z}\in \mathcal{H}_{\mathbf{Z}}} m_t(\boldsymbol{\eta}_{\mathbf{Z}})=f_t(\mathbf{Z})+g_{\mathbf{Z}}\left(\text{grad}f_t({\mathbf{Z}}),\boldsymbol{\eta}_{\mathbf{Z}}\right)
   \nonumber\\
   \!\!\!\!&&\!\!\!\!~~~~~~~~~~~~~~~~~~~~+  \frac{1}{2}g_{\mathbf{Z}}\left(\text{Hess}f_t[\boldsymbol{\eta}_{\mathbf{Z}}],\boldsymbol{\eta}_{\mathbf{Z}}\right)
\nonumber\\
  \!\!\!\!&&\!\!\!\! \textrm{s.t.}~~~
\left \|  \boldsymbol{\eta}_{\mathbf{Z}}\right \|_{g}\leq \Delta_t ,
\end{eqnarray}
where $\left \|  \boldsymbol{\eta}_{\mathbf{Z}}\right \|_{g}=\sqrt{g_{\mathbf{Z}}(\boldsymbol{\eta}_{\mathbf{Z}},\boldsymbol{\eta}_{\mathbf{Z}})}$ and $\Delta_t$ is the trust-region radius.
In this paper, we adapt the Steihaug-Toint
truncated conjugate-gradient (tCG) method from \cite{pa} to approximately solve (\ref{trust}), which is summarized in Algorithm 2. Such a step is integral to locate a critical point of $f(\mathbf{Z})$ by the trust-region Newton method, as formally stated in Algorithm 3, where the criterion for choosing the radius of the trust region is given by
\begin{equation}\label{crit}
  \varrho_t= \frac{f(\mathbf{Z}_t)-f(\mathbf{R}_\mathbf{Z}(\boldsymbol{\eta}_{\mathbf{Z}}^t))}{m_t(\mathbf{0})-m_t(\boldsymbol{\eta}_{\mathbf{Z}}^t)}.
\end{equation}

\begin{algorithm}[h]
\caption{Truncated Conjugate Gradient Algorithm for subproblem (\ref{trust}).}
\label{alg2}
\begin{algorithmic}[1]
\STATE \textbf{Input}: $\mathbf{Z}_t$, $\Delta _t$, Parameters $\theta, \kappa>0$.
\STATE \textbf{Initialization}: $\boldsymbol{\eta}_{\mathbf{Z}}^0=0$,$\boldsymbol{\iota}_0=\text{grad}f(\mathbf{Z}_t)$, $\boldsymbol{\delta} _0=-\boldsymbol{\iota}_0$,$j=0$\\
\WHILE{$\left \|  \boldsymbol{\iota}_{j+1}\right \|_{g}>\left \|  \boldsymbol{\iota}_{0}\right \|_{g}\text{min}(\left \|  \boldsymbol{\iota}_{0}\right \|_{g}^{\theta},\kappa)$}
\IF{$g_{\mathbf{Z}}(\boldsymbol{\delta}_j,\text{Hess}f_t
[\boldsymbol{\delta}_j])\leq 0$}
\STATE Compute $\varpi=\text{argmin}~m_t(\boldsymbol{\eta}_{\mathbf{Z}})$ with constraint $\boldsymbol{\eta}_{\mathbf{Z}}=\boldsymbol{\eta}_{\mathbf{Z}}^j+\varpi \boldsymbol{\delta}_j$ and $\left \|  \boldsymbol{\eta}_{\mathbf{Z}}\right \|_{g}=\Delta_t$
\STATE \bf{return} $\boldsymbol{\eta}_{\mathbf{Z}}^t:=\boldsymbol{\eta}_{\mathbf{Z}}$
\ENDIF\\
Set $\boldsymbol{\eta}_{\mathbf{Z}}^{j+1}=\boldsymbol{\eta}_{\mathbf{Z}}^j+\pounds \boldsymbol{\delta}_j$ with $\pounds =\left \|  \boldsymbol{\iota}_{j}\right \|_{g}^2/g_{\mathbf{Z}}(\boldsymbol{\delta}_j,\text{Hess}f_t
[\boldsymbol{\delta}_j])$
\IF{$\left \|  \boldsymbol{\eta}_{\mathbf{Z}}^{j+1}\right \|_{g}\geq \Delta_t$}
\STATE Compute $\varpi$ as the solution to $\left \|  \boldsymbol{\eta}_{\mathbf{Z}}\right \|_{g}= \Delta_t$ with $\boldsymbol{\eta}_{\mathbf{Z}}=\boldsymbol{\eta}_{\mathbf{Z}}^j+\varpi \boldsymbol{\delta}_j$
\STATE \bf{return} $\boldsymbol{\eta}_{\mathbf{Z}}^t:=\boldsymbol{\eta}_{\mathbf{Z}}$
\ENDIF\\
\STATE $\boldsymbol{\iota}_{j+1}=\boldsymbol{\iota}_{j}+\pounds \text{Hess}f_t
[\boldsymbol{\delta}_j]$
\STATE $\beta_{j+1}=\left \|  \boldsymbol{\iota}_{j+1}\right \|_{g}^2/\left \|  \boldsymbol{\iota}_{j}\right \|_{g}^2$
\STATE $\boldsymbol{\delta}_{j+1}=-\boldsymbol{\iota}_{j+1}+\beta_{j+1}\boldsymbol{\delta}_j$
\STATE $j=j+1$
\ENDWHILE
\STATE \textbf{Output}: $\boldsymbol{\eta}_{\mathbf{Z}}^t$
\end{algorithmic}
\end{algorithm}

\begin{algorithm}[h]
\caption{Signal Recovery for Joint Activity Detection and Channel Estimation.}
\label{alg3}
\begin{algorithmic}[1]
\STATE \textbf{Initialization}: $t\leftarrow 0, \mathbf{Z}_0, \overline{\Delta }\leftarrow \sqrt{r^e}, \Delta_0\leftarrow 0.125\overline{\Delta }, \varrho '\leftarrow 0.1$, tolerance $\varpi_1$\\
\WHILE{$\left \| \text{grad}f(\mathbf{Z}_t) \right \|_F\geq \varpi_1$}
\STATE Obtain $\boldsymbol{\eta}_{\mathbf{Z}}^t$ by approximately solving problem (\ref{trust})
via Algorithm 2.
\STATE Set $\varrho_t$ according to Eq. (\ref{crit})
\STATE If $\varrho_t\leq 0.25$, set $\Delta _{t+1}\leftarrow 0.25\Delta _t$.
\STATE If $\varrho_t\geq  0.75$ and $\left \|  \boldsymbol{\eta}_{\mathbf{Z}}^t\right \|_{g}=\Delta_t$,\\
set $\Delta _{t+1}\leftarrow \text{min}\left \{ 2\Delta_t,\overline{\Delta}  \right \}$;\\
otherwise set $\Delta _{t+1}\leftarrow \Delta _t$.
\STATE If $\varrho_t>\varrho'$, set $\mathbf{Z}_{t+1}=\mathbf{Z}_{t}+\alpha_t\boldsymbol{\eta}_{\mathbf{Z}}^t$;\\
otherwise set $\mathbf{Z}_{t+1}=\mathbf{Z}_{t}$.
\STATE Update $t \leftarrow t+1$
\ENDWHILE
\STATE \textbf{Output}: $\hat{\mathbf{S}}=\mathbf{\overline{P}}\hat{\mathbf{Z}}(\hat{\mathbf{Z}})^H\mathbf{\widetilde{P}}$ with $\hat{\mathbf{Z}}=\mathbf{Z}_{t}$.
\end{algorithmic}
\end{algorithm}

\emph{Remark 4}: Literature \cite{conv} has shown that the Riemannian trust-region algorithm is globally convergent with superlinear convergence rate, i.e. it converges to the second-order KKT points starting from any random initialization. Since the objective function $f(\mathbf{Z})$ is exactly quadratic function which satisfies the Lipschitz gradient condition in \cite{conv}, thus an approximate second-order critical point can always be found by Algorithm 3.

\subsection{Computational Complexity Analysis}
In what follows, we analyze the computational complexity of the proposed algorithm.
\begin{enumerate}

\item The objective function: the complexity of $f(\mathbf{Z})$ in (\ref{lift}) mainly comes from the matrix multiplication, which is in the order of $\mathcal{O}(LNr^e)$ per iteration.

\item The Riemannian gradient: the complexity of Riemannian gradient includes the computation of horizontal projection $\Pi_{\mathbf{Z}}^h$ introduced in (\ref{pro1}), i.e. $\mathcal{O}(N(r^e)^2+(r^e)^3)$, and the computation of matrix multiplication in $\mathcal{O}(LNr^e)$. Thus, the total computational complexity of $\text{grad}f({\mathbf{Z}})$ in (\ref{liftgrad}) is $\mathcal{O}(LNr^e+N(r^e)^2+(r^e)^3)$.

\item The Riemannian Hessian: the overall complexity of $\text{Hess}f(\mathbf{Z})[\boldsymbol{\eta}_{\mathbf{Z}}]$ in (\ref{hes}) is $\mathcal{O}(LNr^e+N(r^e)^2+(r^e)^3)$.

\item The Riemannian metric: the computational complexity of Riemannian metric ${g}_{\mathbf{Z}}(\cdot)$ in (\ref{inner}) is dominated by the matrix multiplication, the complexity of which is $\mathcal{O}(N(r^e)^2)$.

\item The retraction: the computational complexity of retraction introduced in (\ref{sta}) is $\mathcal{O}(Nr^e)$.

\end{enumerate}

It follows from the analysis above that the overall computational complexity of manifold-related operations for solving JADCE problem is $\mathcal{O}(LNr^e+N(r^e)^2+(r^e)^3)$ at each iteration, which does not grow by increasing the number $M$ of BS antennas. Before implementing Riemannian trust-region algorithm, an $r^e \times r^e$ SVD and the rank estimation are computed, which are not numerically expensive, since $r^e$ is less than or equal to the number of active devices. In summary, the proposed DR-JADCE algorithm is computationally efficiency and works well for B5G cellular IoT with massive access.

\section{Numerical Results}
In this section, we report the results of a detailed numerical study to verify the effectiveness of the proposed DR-JADCE algorithm. We first describe the results of experiment that estimate the rank of the interested signal $\mathbf{X}$ with the additive noise. We simulate the underdetermined B5G cellular IoT network with $N = 300$ devices, the actual rank $r^e=30$, a pilot sequence length $L=90$, the unspecified constant $u=0.6+1.2\sqrt{M/L}-1.2M\ln(1+\sqrt{L/M})/L$, and $\beta$ estimated from the proposed in \cite{bet}, c.f. (32) and (33). When measuring rank estimation quality, we consider the mean value of the rank estimation and probability of the successful recovery of the form Pr$(r=r^e)$. Each simulation is repeated 500 times.

\begin{figure}[h]
  \centering
\includegraphics [width=0.5\textwidth] {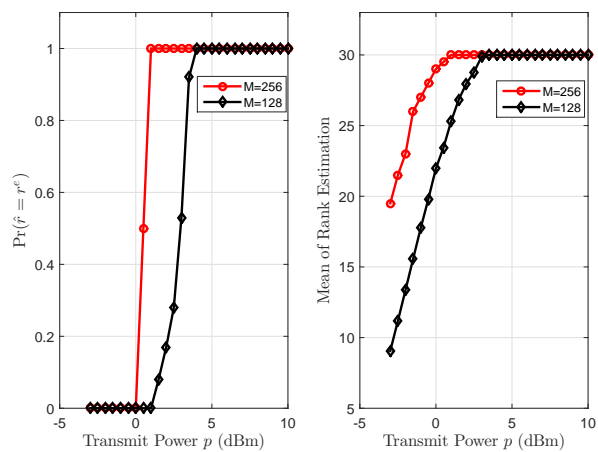}
\caption{The performance of rank estimation algorithm.}
\label{rankestimate}
\end{figure}

Fig. \ref{rankestimate} suggests that, for $M=256$, when the pilot transmit power $-3\leq p<1$ dBm, the success rate is less than 1 and the mean value of rank estimation is increasing from 19.5 to 29.5. However, when $p\geq 1$ dBm, it is sufficient to successfully estimate the rank of interested signal $\mathbf{X}$. For $M=128$, a cut-off point for accurate rank estimation is $p\geq3.5$ dBm. Moreover, the success rate is increased as the antennas number $M$ grows in the lower pilot transmit power region.

Then, we present the device detection performance and channel estimation accuracy of the proposed algorithm. As a reference, we compare the DR-JADCE algorithm with the original $l_{21}$ minimization in (\ref{mmv}), the covariance based detection schemes, i.e. NNLS estimator, MMV estimator and ML estimator \cite{scaling}, whose solution depends on the received signal through certain covariance matrix only, simultaneous orthogonal matching pursuit (SOMP) \cite{OMP1} algorithm, which enhances the detection performance by accumulating the correlation for a group of symbols, AMP algorithm in MMV form \cite{AMPO} which leverages large-scale fading coefficients and the statistics of the wireless channel to improve the estimation performance, and the oracle MMSE algorithm which assumes the support set of the device state matrix is known.

As a performance measure, we use the activity error rate (AER) and normalized mean square error (NMSE) as well. The AER includes missed detection probability, defined as the probability that a device is active but the detector declares it to be inactive, and the false-alarm probability, defined as the probability that a device is inactive, but the detector declares it to be active. The NMSE of the estimated channels of active devices is defined as $10\log_{10}\frac{\left \| \hat{\mathbf{X}}_{\mathcal{K}}-\mathbf{X}_{\mathcal{K}} \right \|^2}{\left \| \mathbf{X}_{\mathcal{K}} \right \|^2}$ where $\mathbf{X}_{\mathcal{K}}$ is for collecting the row-vectors corresponding to active support $\mathcal{K}$ in $\mathbf{X}$.  We set the parameter $\theta=1/0.039$ in the logarithmic function, $\zeta=8$. The large-scale fading $\vartheta$ is assumed to be $-123$dB, and the elements of small-scale fading $\tilde{\mathbf{h}}_n$ obeys Gaussian distribution with zero mean and variance $1$. The power spectral density of the AWGN at the BS is set as $-160$ dBm/Hz, and the bandwidth is set as 1 MHz.

\begin{figure}[h]
  \centering
\includegraphics [width=0.5\textwidth] {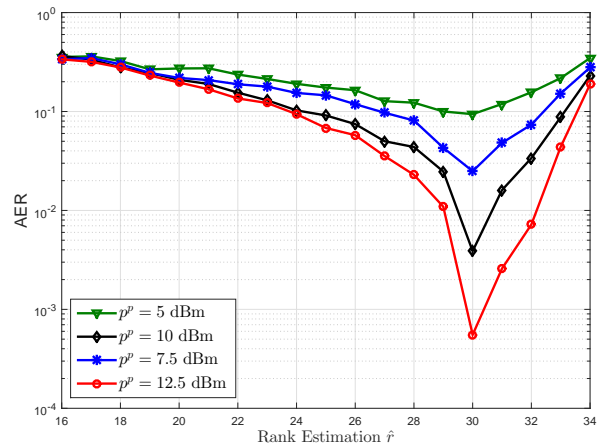}
\caption{The activity error rate for different rank estimation $r$ with a device access probability $\varepsilon=0.1$, devices $N=300$, antennas number $M=64$ and pilot length $L=33$.}
\label{AER_rank}
\end{figure}

\begin{figure}[h]
  \centering
\includegraphics [width=0.51\textwidth] {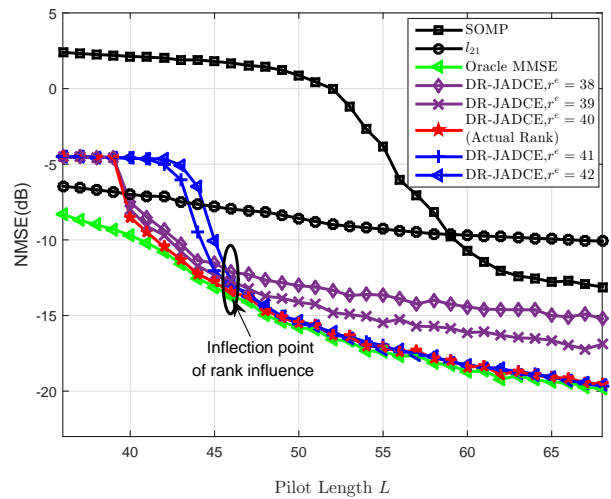}
\caption{The NMSE for different pilot sequence lengths $L$ with a device access
probability $\varepsilon=0.1$, devices $N=400$, pilot transmit power $p=15$ dBm and $M=128$ antennas at the BS.}
\label{ranknmse}
\end{figure}

First, we conduct simulations to validate the effectiveness of the proposed optimization algorithm for activity detection. Fig. \ref{AER_rank} illustrates the evolution of AER when the rank estimation $r$ varies from 16 to 34 and the actual rank is $r^e = 30$. Fig. \ref{AER_rank} shows that the sensitivity of the DR-JADCE to the rank estimation value decreases in low pilot transmit power case compared with ones in high pilot transmit power, which suggests that the proposed algorithm can provide better robustness to rank estimation in the low pilot transmit power regime. Here, an important observation is that the activity error rate of the proposed DR-JADCE algorithm is not sensitive to the accuracy of rank estimation when the rank estimation is less than the actual rank. Although the performance gap between actual rank and overestimated rank in terms of AER is relatively large, the underestimated rank area is what we are really interested in, because even if the actual rank is known, we can utilize a small rank in the proposed
algorithm to further reduce the computational complexity. In addition, as discussed in more detail later in Fig. \ref{ranknmse}, the influence of rank estimation on signal recovery error has an inflection point along the pilot length, which motivates us to adopt different rank setting strategies to reduce the impact of this gap.

\begin{figure}[h]
  \centering
\includegraphics [width=0.5\textwidth] {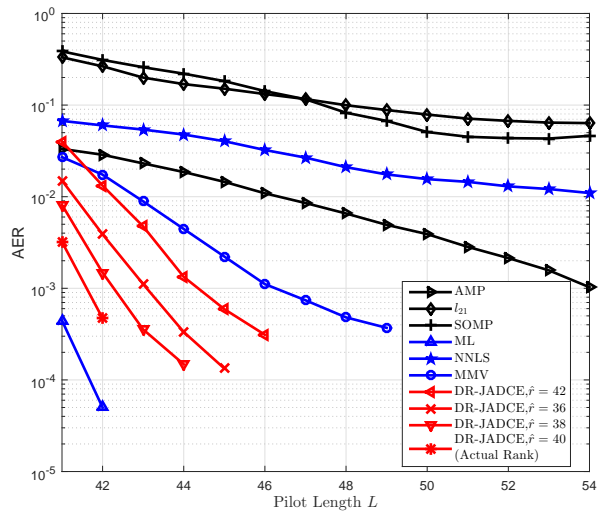}
\caption{The activity error rate for different pilot sequence lengths $L$ with a device access probability $\varepsilon=0.1$, devices $N=400$, pilot transmit power $p=20$ dBm and $M=64$ antennas at the BS.}
\label{AER_pilot}
\end{figure}

Fig. \ref{AER_pilot} examines the AER over the different length of pilot sequences with various rank estimation $r$. It is seen that the overall activity error rate of the DR-JADCE is lower than the $l_{21}$ minimization, MMV, NNLS, AMP and SOMP algorithms. In other words, the proposed DR-JADCE algorithm needs shorter pilot length than other algorithms to achieve the same detection performance. As quality parameter illustrated in Remark 1, the dimension reduction-based approach not only reduces the computation complexity with large antenna array, but also decreases the required length of pilot sequence since $M$ is replaced by a smaller $r$. More importantly, after the rank estimation and dimension reduction operations, the full rank information of $\mathbf{S}$ is incorporated to form a rank aware algorithm. The proposed Riemannian optimization algorithm well incorporate this non-convex full column rank constraints for efficiently decreasing the search space of the JADCE problem. It can converge to first-order and second-order KKT points on manifolds from arbitrary initial points with a superlinear convergence rate. All of this efficiently prove the accuracy of the recovery solution. As can be observed that the proposed DR-JADCE performs worse than ML estimator in terms of detection accuracy. Notice that the ML covariance-based approach needs to detect device activity first, and then estimate channels based on activity detection results, while the proposed algorithm can detect device activity and estimate channels simultaneously. Moreover, the performance degrades as $L$ decreases because the ratio $L/N$ of the system decreases, which indicates that the system becomes more underdetermined.

\begin{figure}[h]
  \centering
\includegraphics [width=0.48\textwidth] {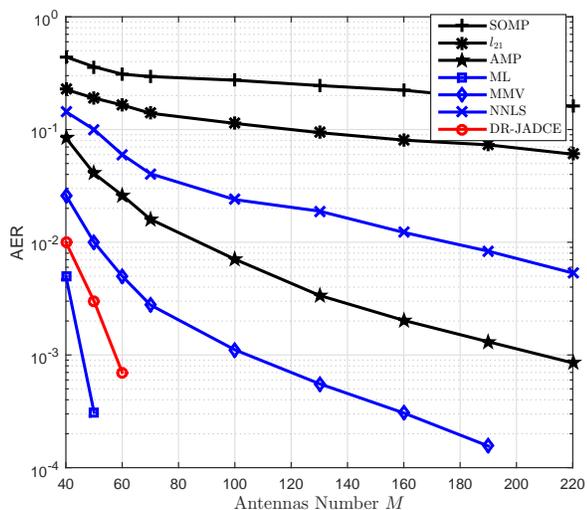}
\caption{The activity error rate for different antennas number $M$ with a device access
probability $\varepsilon=0.1$, devices $N=400$, pilot transmit power $p=15$ dBm and pilot length $L=45$.}
\label{AER_ANTENNA}
\end{figure}

Fig. \ref{AER_ANTENNA} demonstrates the activity detection performance of the DR-JADCE, $l_{21}$ minimization, MMV, NNLS, ML, AMP and SOMP algorithms for different number of antennas with the actual rank setting. It is seen that the DR-JADCE algorithm provides much lower AER than $l_{21}$ minimization, MMV, NNLS, AMP and SOMP algorithms over the number of antennas and the performance gap becomes larger as the number of BS antennas increases. Here, ML performs much better than all the other algorithms and requires much less number of antennas $M$. We emphasize that this paper considers the massive MIMO regime. The reason for this setting is that when the BS is equipped with large antenna array in B5G cellular IoT, the reduction of computational complexity of DR-JADCE is substantial compared with the original one. In the small BS antenna array case, where BS antennas $M$ goes to 1, we would have obtained a low AER for the proposed algorithm, which is not of practical interest. Above mentioned means that superiority of the proposed RD-JADCE scheme is evident for a massive MIMO system, which is widely assumed in the current mMTC.

\begin{figure}[h]
  \centering
\includegraphics [width=0.48\textwidth] {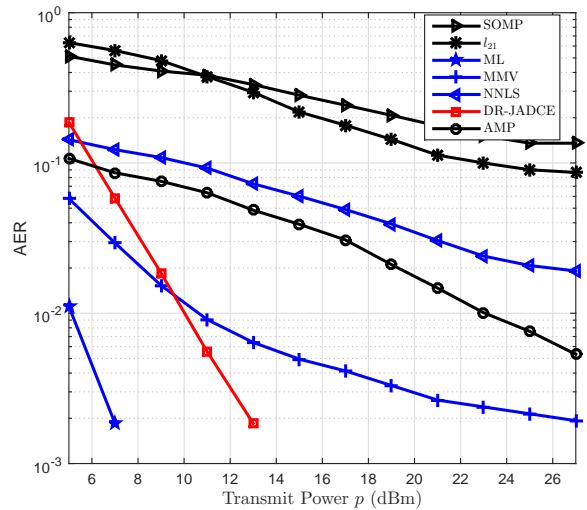}
\caption{The activity error rate for different pilot transmit power $p$ with a device access probability $\varepsilon=0.1$, devices $N=400$, $M=64$ antennas at the BS and pilot length $L=45$.}
\label{AER_SNR}
\end{figure}

Fig. \ref{AER_SNR} plots activity error rate of the three algorithms versus different pilot transmit power with the actual rank setting. We observe that for considered pilot transmit power, the SOMP and $l_{21}$ minimization perform worse than the DR-JADCE algorithm. For $p<6 dBm$, NNLS and AMP algorithms perform better than DR-JADCE algorithm. For $p<9 dBm$, MMV algorithm perform better than the DR-JADCE algorithm. However, the detection performance of DR-JADCE algorithm significantly improves when pilot transmit power increase,
which shows that the proposed algorithm is appealing for mMTC with limited interference.

\begin{figure}[h]
  \centering
\includegraphics [width=0.48\textwidth] {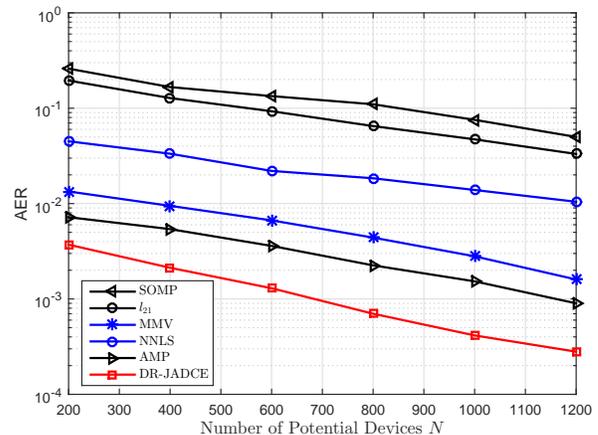}
\caption{The activity error rate for different number of potential devices with active devices $K=100$, transmit power $p=15$ dBm, $M=128$ antennas at the BS and pilot length $L=105$.}
\label{totaldevice}
\end{figure}

Fig. \ref{totaldevice} depicts the AER performance as a function of total number of devices. The total number of devices varies from 200 to 1200. We set the number of active devices $K=100$, the number of antennas $M=128$ and pilot length $L=105$. We can intuitively see that the performance of proposed DR-JADCE algorithm and other compared algorithms are not sensitive to the number of total devices, and the proposed DR-JADCE can provide substantially better AER performance than the $l_{21}$ minimization, MMV, NNLS, AMP and SOMP algorithms over the whole total number of devices range.

\begin{figure}[h]
  \centering
\includegraphics [width=0.51\textwidth] {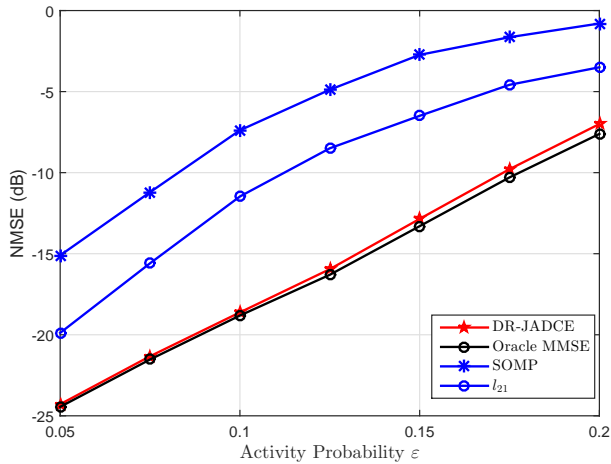}
\caption{The NMSE for different activity probability $\varepsilon$ with pilot transmit power $p=20$ dBm, devices $N=400$, a pilot length $L=90$ and $M=128$ antennas at the BS.}
\label{activity}
\end{figure}

To further illustrate the performance of the proposed
method, we focus on investigating channel estimation accuracy of the proposed algorithm. Fig. \ref{activity} shows the channel estimation performance when the activity probability $\varepsilon$ varies from $0.05$ to $0.2$. We observe that the proposed algorithm outperforms SOMP, $l_{21}$ over the entire activity probability range, and the proposed algorithm and oracle MMSE have similar performance, which indicates that the proposed DR-JADCE algorithm can accommodate more active devices. Furthermore, the performance of all algorithms is degraded as $\varepsilon$ increases. This is because the interference among devices increases as more devices are active.

Fig. \ref{ranknmse} shows the evolution of NMSE of DR-JADCE under various rank estimation value with the actual rank $r^e = 40$. From this figure, we observe that the channel estimation performance increases as pilot length increases, and the DR-JADCE algorithm achieves a substantial performance
gain over the $l_{21}$ minimization, SOMP algorithms. We can also seen that the performance gap between the proposed method and the Oracle MMSE is small, especially when the pilot is longer. This is because the DR-JADCE exploits a full column rank information via an efficient approach. Remarkably, the proposed DR-JADCE algorithm reduces the minimum length of the pilot for stable recovery the device state matrix.

The effect of rank estimation value is clearly observed as we move away from underestimation to overestimation of the rank value. It is seen that underestimating the rank has less influence on the CSI estimation error than overestimating the rank in short pilot length region. However, as $L$ further increases, the performance gap between DR-JADCE with true rank and the case with overestimating rank is dramatically diminished.
In Fig. \ref{ranknmse}, we mark this inflection point (around the number of active devices $+$ $6$) of the influence of the rank estimation on the signal recovery error. In the following, we explain this phenomenon. It is known that the signal-to-interference-plus-noise ratio (SINR) of uplink channel estimation is proportional to the pilot length. In the long pilot region, the high SINR suggests that the signal strength is dominant, so even if the rank estimation is larger than the actual value, the signal can be recovered accurately in the long pilot region. Conversely, underestimating the rank leads to a relatively large estimation error. In the short pilot region, the low SINR suggests that noise has a greater impact on the estimation accuracy, and overestimating the rank motivates more noise to be included. Hence, underestimating the rank leads to a lower NMSE than overestimating the rank in the short pilot region. Hence, underestimating the rank leads to lower NMSE than overestimating the rank in the short pilot region. The Fig. \ref{AER_rank} and Fig. \ref{AER_pilot} are drawn in the short pilot length region, and it seems that if the estimated rank is more than actual rank, the activity detection performance is degraded more severely.

Further, in order to achieve reasonably accurate uplink channel estimation, pilot length $L$ needs to be larger than the number of active devices $K$ in practice. Therefore, in the short pilot region ($K \leq L <K+6$), it is mostly beneficial to further reduce the computation complexity of joint activity detection and channel estimation in DR-JADCE algorithm by taking a small rank even if the actual rank is known. In the long pilot region ($K+6 \leq L$), it is beneficial to guarantee activity detection and channel estimation accuracy of DR-JADCE algorithm by taking a large estimation value of the rank.

\section{Conclusion}
The problem of joint activity detection and channel estimation was studied in
this paper. A dimension reduction model was proposed
by leveraging the low-rank structure of the received data matrix, in
which the interested matrix is full column rank. Based on this model, we developed an optimized design framework with a logarithmic smoothing objective function and a coupled full column rank constraint. To reduce the computational complexity and achieve good
performance, we develop a Riemannian trust region algorithm to solve the rank-constrained smoothed optimization problem by exploiting the complex compact Stiefel manifold of complex fixed-rank matrices. Simulation results shown that the proposed method offers competitive performance in terms of device detection and channel estimation.

\begin{appendices}
\section{The Proof of Proposition 1}
According to the Theorem 2 in \cite{awareness}, if the pilot matrix $\mathbf{A}$: $\Sigma_{K}\rightarrow \mathbb{C}^{L\times M} $ is injective, we have $K\leq \frac{\text{spark}(\mathbf{A})-1+\text{rank}(\mathbf{X})}{2}$,
where $\text{spark}(\mathbf{A})$ denotes the smallest number of linearly dependent columns of $\mathbf{A}$. Then we have
\begin{equation}\label{sbs}
  K\leq \text{spark}(\mathbf{A})-1,
\end{equation}
which comes from the observation that $\text{rank}(\mathbf{X})\leq K$. It follows that
\begin{equation}\label{sbs1}
  K< \text{spark}(\mathbf{A}).
\end{equation}

Naturally we have that $\text{rank}(\mathbf{AX})\leq \text{rank}(\mathbf{X})$. If we assume that $\text{rank}(\mathbf{AX})< \text{rank}(\mathbf{X})$, then, we can obtain a $K$-sparse vector $\mathbf{x}\neq0$ in the span of columns of $\mathbf{X}$ such that $\mathbf{Ax} = \mathbf{0}$, which means that there exists a non-trivial combination of $K$ columns of $\mathbf{A}$ that is equal to 0,  contradicting the inequality that $ K< \text{spark}(\mathbf{A})$ in (\ref{sbs1}). Therefore, $\text{rank}(\mathbf{AX})= \text{rank}(\mathbf{X})$.

\section{The Proof of Lemma 1}
First of all, we rewrite the first term of Eq. (\ref{esre}) by the following transformation
\begin{eqnarray}\label{noi}
&&-(L-r)\ln\left(\frac{\sum_{i=r+1}^{L}\overline{\lambda}_i}{L-r}\right)\nonumber \\
&&=-(L-r)\ln\left(\left(\frac{\sum_{i=r+1}^{L}\overline{\lambda}_i}{L-r}-\sigma^2\right)+\sigma^2\right) \nonumber\\
&&=-(L-r)\left(\left(\frac{\sum_{i=r+1}^{L}\overline{\lambda}_i}{L-r}-\sigma^2\right)\right.\nonumber\\
&&\left.-0.5\left(\frac{\sum_{i=r+1}^{L}\overline{\lambda}_i}{L-r}-\sigma^2\right)^2\left(\sigma^2+\mathcal{O}(\sigma^2)\right)\right)  \nonumber\\
&&=-\sum_{i=r+1}^{L}\left(\overline{\lambda}_i-\sigma^2\right)+\mathcal{O}(\sigma^2),
\end{eqnarray}
where the last equation use $(L-r)({\sigma}^2-\hat{\sigma}^2)^2=\mathcal{O}(1/M)$.
Then define $\bar{\mathcal{L}}_r=-\sum_{i=r+1}^{L}\left(\overline{\lambda}_i-\sigma^2\right)-\sum _{i=1}^r\ln\overline{\lambda}_i$.
When $r<r^e$, we obtain
\begin{eqnarray}\label{noi1}
CM(r^e)-CM(r)&=&\bar{\mathcal{L}}_{r^e}-\bar{\mathcal{L}}_r-\frac{u}{M}(r^e-r)\nonumber \\
&&(L-(r^e+r)/2+0.5).
\end{eqnarray}
According to the result in \cite{bai}, for $r<i<r^e$, we have $\overline{\lambda}_i\rightarrow \lambda_i+\frac{\varrho\lambda_i}{\lambda_i-1}$.
If $\lambda_{r^e}+\frac{\varrho\lambda_{r^e}}{\lambda_{r^e}-1}-\sigma^2-\ln(\lambda_{r^e}+\frac{\varrho\lambda_{r^e}}{\lambda_{r^e}-1})>u\varrho$,
we get
\begin{eqnarray}\label{noi2}
\!\!\!\!\!\!&&\!\!\!\!\!\!P_r\left \{CM(r^e)>CM(r)\right\} \nonumber\\
\!\!\!\!\!\!&&\!\!\!\!\!\!=P_r\left \{\bar{\mathcal{L}}_r^e-\bar{\mathcal{L}}_r>\frac{u}{M}(r^e-r)(L-(r^e+r)/2+0.5) \right \}\nonumber\\
\!\!\!\!\!\!&&\!\!\!\!\!\!=P_r\left \{\sum _{i=r+1}^{r^e}\left(\overline{\lambda}_i-\sigma^2-\ln\overline{\lambda}_i\right) \right.\nonumber \\
\!\!\!\!\!\!&&\!\!\!\!\!\!~~~\left. >\frac{u}{M}(r^e-r)(L-(r^e+r)/2+0.5) \right \}\nonumber\\
\!\!\!\!\!\!&&\!\!\!\!\!\!=P_r\left \{\sum _{i=r+1}^{r^e}\left(\lambda_i+\frac{\varrho\lambda_i}{\lambda_i-1}-\sigma^2-\ln\left(\lambda_i+\frac{\varrho\lambda_i}{\lambda_i-1}\right)\right)\right.\nonumber \\
\!\!\!\!\!\!&&\!\!\!\!\!\!~~~\left.>\frac{u}{M}(r^e-r)(L-(r^e+r)/2+0.5) \right \}\nonumber\\
\!\!\!\!\!\!&&\!\!\!\!\!\! \geq P_r\left \{\lambda_{r^e}+\frac{\varrho\lambda_{r^e}}{\lambda_{r^e}-1}-\sigma^2-\ln\left(\lambda_{r^e}+\frac{\varrho\lambda_{r^e}}{\lambda_{r^e}-1}\right)\right.\nonumber \\
\!\!\!\!\!\!&&\!\!\!\!\!\!~~~\left.>\frac{u}{M}(L-\frac{r^e-1}{2}))\right \} \rightarrow 1.
\end{eqnarray}
When $r>r^e$, combining the result in \cite{bai}, i.e. for $r^e<i<r$, $\overline{\lambda}_i\rightarrow (1+\sqrt{\varrho})^2$, we can write
\begin{eqnarray}\label{noi3}
\!\!\!\!\!\!&&\!\!\!\!\!\!P_r(CM(r^e)>CM(r)) \nonumber\\
\!\!\!\!\!\!&&\!\!\!\!\!\!=P_r\left \{\sum _{i=r+1}^{r^e}\left(\overline{\lambda}_i-\sigma^2-\ln\overline{\lambda}_i\right)\right.\nonumber \\
\!\!\!\!\!\!&&\!\!\!\!\!\!~~~\left.>\frac{u}{M}(r^e-r)(L-(r^e+r)/2+0.5) \right \}\nonumber \\
\!\!\!\!\!\!&&\!\!\!\!\!\!=P_r\left \{(1+\sqrt{\varrho})^2-\sigma^2-2\ln(1+\sqrt{\varrho})\right.\nonumber \\
\!\!\!\!\!\!&&\!\!\!\!\!\!~~~\left.>\frac{u}{M}(r^e-r)(L-(r^e+r)/2+0.5) \right \}\nonumber \\
\!\!\!\!\!\!&&\!\!\!\!\!\!=P_r\left \{ u>\frac{L}{L-(r^e+r)/2+0.5}\frac{M}{L} \right.\nonumber \\
\!\!\!\!\!\!&&\!\!\!\!\!\!~~~\left.\left(1-\sigma^2+\varrho+2\sqrt{\varrho}-2\ln(1+\sqrt{\varrho})\right)\right\}\rightarrow 1,
\end{eqnarray}
if $u>\frac{1-\sigma^2}{\varrho}+1+2\sqrt{1/\varrho}-2\ln(1+\sqrt{\varrho})/\varrho$ holds. This completes the proof.

\section{The Proof of the smoothness of the proposed logarithmic smooth method}
Now, we prove that the method in (\ref{logr}) can solve the nonsmooth problem. $\mathbf{z}=J(\mathbf{x})$ is differentiable at $\mathbf{0}$ if the following condition holds
\begin{eqnarray}\label{difffer}
\!\!\!\!\!\!&&\!\!\!\!\!\!\Delta\mathbf{z}-J_{\mathbf{x}_1}(\mathbf{0})\Delta \mathbf{x}_1-J_{\mathbf{x}_2}(\mathbf{0})\Delta \mathbf{x}_2-\cdots-J_{\mathbf{x}_{r^e}}(\mathbf{0})\Delta \mathbf{x}_{r^e} \nonumber \\
\!\!\!\!\!\!&=&\!\!\!\!\!\!\mathcal{O}\left(\sqrt{\Delta \mathbf{x}_1^2+\Delta \mathbf{x}_2^2+\cdots+\Delta \mathbf{x}_{r^e}^2}\right),
\end{eqnarray}
where $\mathbf{x}_{i}$ denotes the $i$-th element of the vector $\mathbf{x}$, $\Delta \mathbf{x}_{i}$ denotes the micro change of element $\mathbf{x}_{i}$. $J_{\mathbf{x}_i}(\mathbf{0})$ represents the value of the partial derivative of $J$ to $\mathbf{x}_i$ at point $\mathbf{0}$. Calculating the partial derivative
\begin{eqnarray}\label{difffer1}
J_{\mathbf{x}_i}(\mathbf{0}^+)=J_{\mathbf{x}_i}(\mathbf{0}^-)\!\!\!\!\!\!&=&\!\!\!\!\!\!\lim_{\mathbf{x}_i\rightarrow 0}\frac{\left \| \mathbf{x}_i \right \|_2-\frac{1}{\theta}\ln(1+\theta\left \| \mathbf{x}_i \right \|_2)}{\mathbf{x}_i} \nonumber\\
\!\!\!\!\!\!&=&\!\!\!\!\!\!\lim_{\mathbf{x}_i\rightarrow 0}\frac{\left \| \mathbf{x}_i \right \|_2-\frac{1}{\theta}\theta\left \| \mathbf{x}_i \right \|_2}{\mathbf{x}_i}=0,
\end{eqnarray}
and combining the Maclaurin series of logarithmic function, Eq. (\ref{difffer}) reduces to
\begin{eqnarray}\label{difffer2}
\Delta \mathbf{z}\!\!\!\!\!\!&=&\!\!\!\!\!\!\sqrt{\Delta \mathbf{x}_1^2+\Delta \mathbf{x}_2^2+\cdots+\Delta \mathbf{x}_{r^e}^2}\nonumber \\
\!\!\!\!\!\!&&\!\!\!\!\!\!-\frac{1}{\theta}\ln\left(1+\theta\sqrt{\Delta \mathbf{x}_1^2+\Delta \mathbf{x}_2^2+\cdots+\Delta \mathbf{x}_{r^e}^2}\right)\nonumber\\
\!\!\!\!\!\!&=&\!\!\!\!\!\!\sqrt{\Delta \mathbf{x}_1^2+\Delta \mathbf{x}_2^2+\cdots+\Delta \mathbf{x}_{r^e}^2}\nonumber \\
\!\!\!\!\!\!&&\!\!\!\!\!\!-\frac{1}{\theta }\left(\theta \sqrt{\Delta \mathbf{x}_1^2+\Delta \mathbf{x}_2^2+\cdots+\Delta \mathbf{x}_{r^e}^2} \nonumber \right. \\
\!\!\!\!\!\!&&\!\!\!\!\!\! \left.-\frac{\theta^2}{2}(\Delta \mathbf{x}_1^2+\Delta \mathbf{x}_2^2+\cdots+\Delta \mathbf{x}_{r^e}^2)
+\cdots \right)\nonumber\\
\!\!\!\!\!\!&=&\!\!\!\!\!\!\mathcal{O}\left(\sqrt{\Delta \mathbf{x}_1^2+\Delta \mathbf{x}_2^2+\cdots+\Delta \mathbf{x}_{r^e}^2}\right).
\end{eqnarray}
Therefore, $J( \mathbf{x})$ is differentiable and (\ref{logr}) is a valid smoothing method for $\left\| \mathbf{x}\right\|_{2}$.

\section{The Proof of Vertical Space, Horizontal Space and Horizontal Projection.}
The vertical space at $\mathbf{Z}$ is by definition the tangent space to the equivalence class $[\mathbf{Z}]$. Define $\mathbf{Z}(t)=\mathbf{Z}_0\mathbf{Q}(t)$ as a curve in $[\mathbf{Z_0}]$ through $\mathbf{Z_0}$ at $t=0$, i.e. $\mathbf{Q}(0)=\mathbf{I}$. It follows that
\begin{equation}\label{squ}
  \mathbf{Z}(t)\mathbf{Z}(t)^H=\mathbf{Z}_0\mathbf{Q}(t)\mathbf{Q}(t)^H\mathbf{Z}_0^H=\mathbf{Z}_0\mathbf{Z}_0^H,
\end{equation}
for all $t$.
Differentiating Eq.(\ref{squ}) with respect to $t$ yields
\begin{equation}\label{DIFF}
  \dot{\mathbf{Z}(t)}\mathbf{Z}(t)^H+\mathbf{Z}(t)\dot{\mathbf{Z}(t)}^H=0.
\end{equation}
Therefore, $\dot{\mathbf{Z}(0)}$ is an element of the set
\begin{equation}\label{seet}
  \{\hat{{\mathbf{Z}}}\in \mathbb{C}^{(N+r^e)\times r^e} :\hat{\mathbf{Z}}{\mathbf{Z}}_0^H+ \mathbf{Z}_0\hat{\mathbf{Z}}^H=0\}.
\end{equation}
According to Section 3.5.7 in \cite{pa}, we obtain
\begin{eqnarray}\label{ker}
   \mathcal{T}_\mathbf{Z}{\mathcal{M}}&=&\text{ker}\left(Df_1(\mathbf{Z}_{0})\right)\nonumber\\
   &=&\{\hat{{\mathbf{Z}}}\in \mathbb{C}^{(N+r^e)\times r^e} :\hat{\mathbf{Z}}{\mathbf{Z}}_0^H+ \mathbf{Z}_0\hat{\mathbf{Z}}^H=0\},
\end{eqnarray}
for the quotient space $\mathcal{M}$ defined by function $f_1:\mathbf{Z} \mapsto \mathbf{Z}\mathbf{Z}^H$, where the kernel ker($\mathbf{Z}$) of a matrix $\mathbf{Z}$ is the subspace formed by the vectors $\mathbf{x}$
such that $\mathbf{Zx} = 0$. Since $\mathbf{Z}(t) \in \mathcal{M}$ is full column rank, $\dot{\mathbf{Z}(t)}$ can be set as
\begin{equation}\label{dot}
  \dot{\mathbf{Z}(t)}= \mathbf{Z}(t)\mathbf{B}(t)): \mathbf{B} (t)\in \mathbb{C}^{r^e\times r^e}.
\end{equation}
Substituting Eq. (\ref{dot}) into Eq. (\ref{DIFF}), we finally get (\ref{vers}).

As we show in Eq. (\ref{inner}), it follows that
\begin{eqnarray}\label{HOF1}
     g_{\mathbf{Z}}(\boldsymbol{\xi}_{\mathbf{Z}},\mathbf{Z}\mathbf{B)}
&=&\frac{1}{2}\text{Tr}\left(\boldsymbol{\xi}_{\mathbf{Z}}^H\mathbf{Z}\mathbf{B}+\mathbf{B}^H\mathbf{Z}^H\boldsymbol{\xi}_{\mathbf{Z}}\right)\nonumber\\
&=&\frac{1}{2}\text{Tr}\left(\left(\boldsymbol{\xi}_{\mathbf{Z}}^H\mathbf{Z}-\mathbf{Z}^H\boldsymbol{\xi}_{\mathbf{Z}}\right)\mathbf{B}\right).
\end{eqnarray}
According to the definition in Eq. (\ref{HOF}), the horizontal space is given by
\begin{equation}\label{HS1}
   \mathcal{H}_{\mathbf{Z}}=\left \{ \boldsymbol{\xi}_{\mathbf{Z}}\in \mathbb{C}^{(N+r^e)\times r^e}: \boldsymbol{\xi}_{\mathbf{Z}}^H\mathbf{Z}=\mathbf{Z}^H\boldsymbol{\xi}_{\mathbf{Z}}\right \}.
\end{equation}
Assume that the vertical projection of a vector $\overline{\boldsymbol{\xi}}_{\mathbf{Z}}\in \mathcal{T}_\mathbf{Z}\mathcal{M}$ is given by $\mathbf{Z}\mathbf{B}$. Accordingly, the horizontal projection is accomplished with the operator
\begin{equation}\label{sd}
  \boldsymbol{\xi}_{\mathbf{Z}}=\overline{\boldsymbol{\xi}}_{\mathbf{Z}}-\mathbf{Z}\mathbf{B}.
\end{equation}
By substituting (\ref{sd}) into (\ref{HS1}), we have the coupled system of Lyapunov equation expressed as
\begin{equation}\label{lya}
(\overline{\boldsymbol{\xi}}_{\mathbf{Z}}-\mathbf{Z}\mathbf{B})^H\mathbf{Z}=\mathbf{Z}^H(\overline{\boldsymbol{\xi}}_{\mathbf{Z}}-\mathbf{Z}\mathbf{B}).
\end{equation}
Then decomposition the equation above, we can obtain the result in Eq. (\ref{ome}).
\section{The Proof of Computing the Riemannian Gradient}
According to Eq. (\ref{lift}), the complex gradient of $f(\mathbf{Z})$ with respect to $\mathbf{Z}$ is calculated as
\begin{eqnarray}\label{gradp1}
\!\!\!\!\!\!&&\!\!\!\!\!\!\overline{\text{grad}f(\mathbf{Z})}=\frac{\zeta}{2}\overline{\mathbf{P}}^H\mathbf{A}^H\left(\mathbf{A}\overline{\mathbf{P}}\mathbf{Z}
\mathbf{Z}^H\widetilde{\mathbf{P}}-\mathbf{V}\right)\widetilde{\mathbf{P}}^H\mathbf{Z}\nonumber\\
\!\!\!\!\!\!&&\!\!\!\!\!\!+\frac{\zeta}{2}\widetilde{\mathbf{P}}\left(\mathbf{A}
\overline{\mathbf{P}}\mathbf{Z}\mathbf{Z}^H\widetilde{\mathbf{P}}-\mathbf{V}\right)^H\mathbf{A}\overline{\mathbf{P}}\mathbf{Z}\nonumber \\
\!\!\!\!\!\!&&\!\!\!\!\!\!+\overline{\mathbf{P}}^H\left[ \frac{\left \| \mathbf{A}^H\overline{\mathbf{D}}[1] \right \|_2 \theta\boldsymbol{\Omega }_1^H}{1+\theta\left \| \boldsymbol{\Omega}_1\right \|_2},\cdots,\frac{\left \| \left(\mathbf{A}^H\overline{\mathbf{D}}\right)[N] \right \|_2 \theta\boldsymbol{\Omega }_N^H}{1+\theta\left \| \boldsymbol{\Omega}_N\right \|_2} \right]^H \widetilde{\mathbf{P}}^H\mathbf{Z}\nonumber\\
\!\!\!\!\!\!&&\!\!\!\!\!\!+\widetilde{\mathbf{P}}\left[ \frac{\left \| \mathbf{A}^H\overline{\mathbf{D}}[1] \right \|_2 \theta\boldsymbol{\Omega }_1^H}{1+\theta\left \| \boldsymbol{\Omega}_1\right \|_2},\cdots,\frac{\left \| \left(\mathbf{A}^H\overline{\mathbf{D}}\right)[N] \right \|_2 \theta\boldsymbol{\Omega }_N^H}{1+\theta\left \| \boldsymbol{\Omega}_N\right \|_2} \right]\overline{\mathbf{P}}\mathbf{Z},\nonumber
\end{eqnarray}
The Riemannian gradient is derived from Eq. (\ref{ggrad}), and it is found that
\begin{eqnarray}\label{eqal}
\!\!\!\!\!\!&&\!\!\!\!\!\!Df[\boldsymbol{Z}][\boldsymbol{\xi}_{\mathbf{Z}}]=\frac{1}{2}\left( \sum_{n=1}^{N}\frac{\theta\left(\mathbf{A}^H\overline{\mathbf{D}}\right)[n]}{1+\theta\left \| \boldsymbol{\Omega }_n \right \|_2}\left(\boldsymbol{\Omega }_n\mathbf{\Lambda} _n^H+\mathbf{\Lambda}_n\boldsymbol{\Omega }_n^H\right)\right)\nonumber\\
\!\!\!\!\!\!&&\!\!\!\!\!\!\left.+\frac{\zeta}{2}\text{tr}\left(\left(\mathbf{A}\overline{\mathbf{P}}\mathbf{Z}\boldsymbol{\xi}_{\mathbf{Z}}^H\widetilde{\mathbf{P}}
+\mathbf{A}\overline{\mathbf{P}}\boldsymbol{\xi}_{\mathbf{Z}}\mathbf{Z}^H\widetilde{\mathbf{P}}\right)\left(\mathbf{A}
\overline{\mathbf{P}}\mathbf{Z}\mathbf{Z}^H\widetilde{\mathbf{P}}-\mathbf{V}\right)^H \right.\right.\nonumber\\
\!\!\!\!\!\!&&\!\!\!\!\!\!\left.+\left(\mathbf{A}\overline{\mathbf{P}}\mathbf{Z}\mathbf{Z}^H\widetilde{\mathbf{P}}-\mathbf{V}\right)\left(\widetilde{\mathbf{P}}
^H\boldsymbol{\xi}_{\mathbf{Z}}\mathbf{Z}^H\overline{\mathbf{P}}^H\mathbf{A} +\widetilde{\mathbf{P}}^H\mathbf{Z}\boldsymbol{\xi}_{\mathbf{Z}}^H\overline{\mathbf{P}}^H\mathbf{A}^H\right)\right)\nonumber\\
\!\!\!\!\!\!&&\!\!\!\!\!\!=g_{\mathbf{Z}}(f'(\mathbf{Z}),\boldsymbol{\xi}_{\mathbf{Z}}).
\end{eqnarray}
Correspondingly, we conclude that $\overline{\text{grad}f(\mathbf{Z})}=f'(\mathbf{Z})$.
The proof completes.
\end{appendices}

\end{document}